\documentclass[aps,pre,reprint,groupedaddress]{revtex4-2}

\usepackage{color}
\usepackage{amsmath}
\usepackage{amsfonts} 
\usepackage[pdftex]{graphicx}
\usepackage{booktabs}
\usepackage{url}

\def\mat#1{#1}
\def\bra#1{\mbox{\boldmath $#1$}^{\top}}
\def\ket#1{\mbox{\boldmath $#1$}}

\newcommand{\bracket}[1]{\left\langle #1 \right\rangle}

\def\imgunit{i}

\begin{document}


\title{Spectral clustering of annotated graphs using a factor graph representation}


\author{Tatsuro Kawamoto}
\affiliation{  Artificial Intelligence Research Center, \\
  National Institute of Advanced Industrial Science and Technology, \\
  2-3-26 Aomi, Koto-ku, Tokyo, Japan }



\date{\today}

\begin{abstract}
Graph-structured data commonly have node annotations. 
A popular approach for inference and learning involving annotated graphs is to incorporate annotations into a statistical model or algorithm. 
By contrast, we consider a more direct method named scotch-taping, in which the structural information in a graph and its node annotations are encoded as a factor graph. 
Specifically, we establish the mathematical basis of this method in the spectral framework. 
\end{abstract}


\maketitle

\section{Introduction}
Node annotations (features or attributes) are significantly common in graph datasets. Examples include keywords of papers in citation networks, a person's age and gender in social networks, and group labels of nodes in the form of metadata (occasionally termed as ``ground truth'') in graphs \cite{karateclub,Newman2006politicalbooks} that are used as benchmarks in community detection problems. 
Several methods have been proposed in machine learning and network science for structural inference and learning, or dimensionality reduction, for such data \cite{Wu_GNNreview2019,Zhang_GNNreview2020,chunaev2019community,NewmanClauset2016,Hric_PRX2016}. 
In this study, we focus on discrete node labels that are considered as nominal variables and refer to them as annotations. 
We also restrict the scope to the inference of a module structure, instead of considering a general inference task on annotated graphs. 

A typical approach involves treating a graph as a primary object and incorporating node annotations. 
Examples of this approach are Bayesian inference for graphs, in which node annotations are incorporated as a prior distribution \cite{NewmanClauset2016,Hric_PRX2016}, and constrained-optimization methods \cite{Rangapuram2012,Xiang2014,Peel2017}. 
Another typical approach involves treating node attributes (including ordinal and numerical variables) as primary objects and incorporating the graph structure in a perturbative manner. 
Representative examples of this approach are the frameworks of graph neural networks (GNNs) \cite{Wu_GNNreview2019,Zhang_GNNreview2020,KipfWelling2016,Hamilton_NIPS2017}. 
We note that all the aforementioned methods incorporate node annotations and attributes in a model-dependent and algorithm-dependent manner. 

In this study, we consider a data representation method in which the information contained in a graph and its node annotations are encoded as a factor graph (hypergraph or bipartite graph). 
We refer to this graph as a \textit{scotch-taped graph} and to the representation method as \textit{scotch-taping}. 
We define the scotch-taped graph in the next section and address specific questions. 
In contrast to the methods mentioned above, scotch-taping is based on only the data representation. 
Therefore, we can always consider using the scotch-taped graph as input to an arbitrary algorithm to encode information provided as annotations.

\section{Factor graph representation of a graph with annotated nodes}\label{sec:FactorGraph}
\begin{figure*}
  \centering
  \includegraphics[width= 1.8\columnwidth, bb = 0 0 515 117]{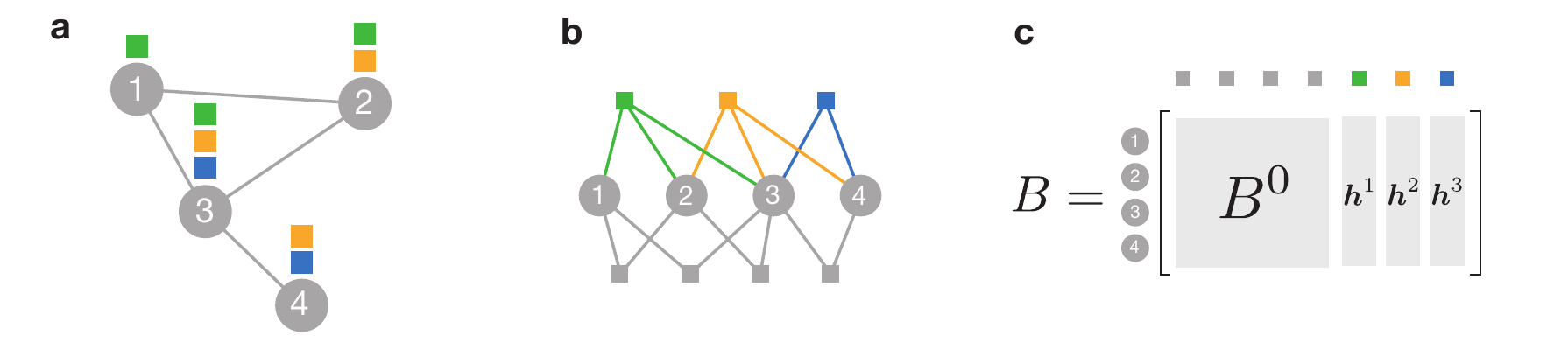}
  \caption{Example of ({\bf a}) a graph $G = (U, E)$ with annotations, ({\bf b}) factor graph representation, i.e., a scotch-taped graph, and ({\bf c}) the corresponding incidence matrix $\mat{B}$. Each colored square represents an annotation; in ({\bf a}), node 1 has only the green annotation, whereas node 2 has the green and yellow annotations, etc.}
  \label{fig:ScotchTaping}
\end{figure*}

As illustrated in Fig.~\ref{fig:ScotchTaping}{\bf a}, we consider a graph $G = (U, E)$ consisting of a node set $U$ ($|U|=N$) and an edge set $E$ ($|E| = M_{0}$). 
We first consider the factor graph representation of $G$. 
It consists of two types of node sets: the physical nodes corresponding to $U$ and the factor nodes $V^{0}$ ($|V^{0}| = M_{0}$) corresponding to the edge set $E$. 
When nodes $i$ and $j$ are connected by an edge in $G$, the factor node $\alpha \in V^{0}$ is connected to $i$ and $j$ in the factor graph. 
We use indices $i$ and $\alpha$ to represent node labels as well as elements of the node sets although this is a slight abuse of notation. 
A set consisting of a factor node and the edges incident to it is termed as a hyperedge in this study. 
The incidence matrix $\mat{B}^{0}$ of the factor graph representation of $G$ is an $N \times M_{0}$ rectangular matrix with elements $B^{0}_{i\alpha}$, where $B^{0}_{i\alpha}=1 $ if an edge exists between the physical node $i$ and the factor node $\alpha$ and $B^{0}_{i\alpha} = 0$ otherwise. 

We introduce an indicator variable $h^{r}_{i} \in \{0,1\}$ that represents whether a physical node $i$ has a certain annotation label $r$ where $r \in \{1, \dots, R\}$. 
We incorporate relationships that the annotations indicate by attaching external hyperedges to the factor graph. 
For example, the $r$th annotation label constitutes an external hyperedge such that a factor node corresponding to the $r$th annotation is connected to a physical node $i$ if $h^{r}_{i} = 1$ (Fig.~\ref{fig:ScotchTaping}{\bf b}). 
Therefore, the overall incidence matrix $\mat{B}$ is defined by the following $N \times M$ ($M = M_{0} + R$) concatenated matrix: 
\begin{align}
\mat{B} = [\mat{B}^{0}, \ket{h}^{1}, \cdots, \ket{h}^{R}] = [\mat{B}^{0}, \mat{H}], \label{TotalIncidenceMatrix}
\end{align}
where $\ket{h}^{r}$ is an $N$-dimensional column vector (Fig.~\ref{fig:ScotchTaping}{\bf c}).  
$\mat{H} = [\ket{h}^{1}, \cdots, \ket{h}^{R}]$ is the concatenated matrix of the external hyperedges. 
The scotch-taped graph is defined as the graph corresponding to this incidence matrix $\mat{B}$. 

We also note that an external hyperedge does not necessarily indicate similarity among the target physical nodes. 
For example, it is possible to let an algorithm learn that the hyperedge indicates a dissimilarity relationship among the target physical nodes. 
Furthermore, we can explicitly label the edges and factor nodes, although this is beyond the scope of the present study. 

An important question that should be addressed is \textit{the effect of scotch-taping on inference}. 
For example, if a graph exhibits a certain module structure and the node annotations exhibit the same structure with a higher resolution (i.e., their combination exhibits a more definite module or hierarchical module structure), we expect that a more detailed inference can be achieved through scotch-taping. 
In contrast, if a graph and its node annotations exhibit qualitatively different structures, they may only act as noise to each other, or the scotch-taped graph may exhibit yet another structure. 
More specifically, let us consider an annotation label that most nodes have. 
Then, almost all physical nodes in the scotch-taped graph are connected to each other through the corresponding factor node. 
It is conceivable that such a single external hyperedge may disrupt the structural information in the original graph. 
To investigate this, we require a systematic understanding of the effect of scotch-taping under certain concrete settings.

We treat a graph as a primary object and incorporate node annotations as a perturbation to the graph. 
In Sec.~\ref{sec:SpectralFramework}, we study the contribution of scotch-taping in the framework of spectral clustering from various perspectives. 
We begin with a formal solution of the eigenvalue equation for a general scotch-taped graph using the Green's function formalism (Sec.~\ref{sec:GreensFunction}). 
Then, focusing on graphs generated by a random graph model, we study the behavior of the leading eigenvalues and eigenvectors; after investigating the extent to which an analysis can be performed using a crude approximation, we derive a mean-field solution that considers more detailed information from a scotch-taped graph (Sec.~\ref{sec:SBMSpectralMethod}). 
Finally, in Sec.~\ref{sec:Discussion}, we briefly discuss the application of scotch-taping to methods other than spectral clustering.

\section{Spectral clustering of scotch-taped graphs}\label{sec:SpectralFramework}
We define the normalized incidence matrix 
\begin{align}
\mat{\mathsf{B}} \equiv \mat{D}^{-1/2}_{U} \mat{B} \mat{D}^{-1/2}_{V}, 
\end{align}
where the degree matrices, $\mat{D}_{U}$ and $\mat{D}_{V}$, are defined as
\begin{align}
\mat{D}_{U} \equiv \mathrm{diag}( d^{u}_{1}, \dots, d^{u}_{N} ) \hspace{10pt} \left(d^{u}_{i} = \sum_{\alpha=1}^{M} B_{i\alpha}\right), \notag\\
\mat{D}_{V} \equiv \mathrm{diag}( d^{v}_{1}, \dots, d^{v}_{M} ) \hspace{10pt} \left(d^{v}_{\alpha} = \sum_{i=1}^{N} B_{i\alpha}\right). 
\end{align}
$\mathrm{diag}(x_{1}, \dots, x_{N})$ represents a diagonal matrix with diagonal elements $x_{1}, \dots, x_{N}$. 
We note that $\mat{D}_{U}$ is affected by external hyperedges; we denote the degree matrix of the original graph by $\mat{D}^{0}_{U}$ and let $\mat{D}^{h}_{U} \equiv \mat{D}_{U} - \mat{D}^{0}_{U} = \mathrm{diag}\left( \sum_{r} h^{r}_{1}, \dots, \sum_{r} h^{r}_{N} \right)$. 
According to spectral graph theory \cite{Dhillon2001}, when a graph has a module structure with $K$ groups, a low-dimensional representation that captures this structure can be obtained by the $K$ leading singular vectors of $\mat{\mathsf{B}}$. 
The $k$th singular value, $s_{k}$, of $\mat{\mathsf{B}}$ satisfies 
\begin{align}
\mat{\mathsf{B}} \ket{\psi}^{\prime}_{k} = s_{k} \ket{\phi}^{\prime}_{k}, 
\hspace{20pt} 
\mat{\mathsf{B}}^{\top} \ket{\phi}^{\prime}_{k} = s_{k} \ket{\psi}^{\prime}_{k}, \label{SingularValueEquation}
\end{align}
where $\top$ denotes the transpose, and $\ket{\psi}^{\prime}_{k}$ and $\ket{\phi}^{\prime}_{k}$ represent the $M$-dimensional right singular vector and $N$-dimensional left singular vector, respectively. 

Hereafter, instead of the pair of the singular-value equations, we consider the equivalent eigenvalue equation with respect to $2\mat{\mathsf{B}}\mat{\mathsf{B}}^{\top}$ with eigenvalue $\lambda_{k} = 2 s^{2}_{k}$. 
This can be transformed into a generalized eigenvalue equation by setting $\ket{\phi}_{k} = \mat{D}^{-1/2}_{U} \ket{\phi}^{\prime}_{k}$. 
By using the internal structure of $\mat{B}$ in $2\mat{\mathsf{B}}\mat{\mathsf{B}}^{\top}$ and rearranging the generalized eigenvalue equation, we can further reformulate Eq.~(\ref{SingularValueEquation}) as 
\begin{align}
2 \mat{D}^{-1/2}_{U} \mat{B} \mat{D}^{-1}_{V} \mat{B}^{\top} \mat{D}^{-1/2}_{U} \ket{\phi}^{\prime}_{k} &= \lambda_{k} \ket{\phi}^{\prime}_{k} \notag\\
2 \mat{B} \mat{D}^{-1}_{V} \mat{B}^{\top} \ket{\phi}_{k} &= \lambda_{k} \mat{D}_{U} \ket{\phi}_{k} \notag\\
2 \left[ \mat{B}^{0}, \ket{h}^{1}, \dots, \ket{h}^{R} \right] \mat{D}^{-1}_{V} \left[ \mat{B}^{0}, \ket{h}^{1}, \dots, \ket{h}^{R} \right]^{\top} 
\ket{\phi}_{k} &= \lambda_{k} \mat{D}_{U} \ket{\phi}_{k} \notag\\
\left( \mat{B}^{0}\mat{B}^{0 \top} - 2\sum_{r=1}^{R} \frac{\ket{h}^{r}\ket{h}^{r \top}}{d^{v}_{r}} \right) \ket{\phi}_{k} 
&= \lambda_{k} \mat{D}_{U} \ket{\phi}_{k}. \label{EigenValueEquation0}
\end{align}
Here, we used the fact that $d^{v}_{\alpha} = 2$ for any $\alpha \in V^{0}$ 
We define the adjacency matrix of the original graph as $\mat{A} = \mat{B}^{0} \mat{B}^{0 \top} - \mat{D}^{0}_{U}$ and introduce the combinatorial Laplacian, $\mat{L} \equiv \mat{D}^{0}_{U} - \mat{A}$. 
Then, Eq.~(\ref{EigenValueEquation0}) can be written as 
\begin{align} 
\left( \mat{L} - 2\sum_{r} \frac{\ket{h}^{r}\ket{h}^{r \top}}{d^{v}_{r}} \right) \ket{\phi}_{k} 
= \left( (2-\lambda_{k}) \mat{D}^{0}_{U} - \lambda_{k} \mat{D}^{h}_{U} \right) \ket{\phi}_{k}. \label{EigenValueEquation}
\end{align}
In the absence of external hyperedges, Eq.~(\ref{EigenValueEquation}) reduces to the generalized eigenvalue equation of $\mat{L}$ with eigenvalue $2-\lambda_{k}$, which is often considered in spectral clustering \cite{Luxburg2007}.
Spectral embedding uses the $K$ leading eigenvectors to obtain a $K$-dimensional representation of each node. 
Spectral clustering is a classification of the result of a low-dimensional spectral embedding into $K$ groups. 
In the case of bipartitioning ($K=2$), the classification of the $i$th (physical) node is often determined based on the sign of the $i$th second eigenvector element \cite{Luxburg2007}. 

We note that eigenvalue $\lambda_{k}$ is nonnegative by definition. 
The largest eigenvalue is $\lambda_{1}=2$, with $\ket{\phi}_{k} \propto \ket{1}_{N}$, where $\ket{1}_{N}$ is an $N$-dimensional column vector with all elements equal to unity; the fact that this is the largest non-degenerate eigenvalue follows from the Perron--Frobenius theorem, assuming that the scotch-taped graph is connected. 
Thus, the eigenvalues of $2\mat{\mathsf{B}}\mat{\mathsf{B}}^{\top}$ are bounded. 

We also note that $2\mat{\mathsf{B}}\mat{\mathsf{B}}^{\top}$ can be regarded as the adjacency matrix of a weighted graph with respect to its physical nodes. The operation to generate such a weighted graph is termed monopartite (or one-mode) projection. 
Therefore, as far as the aforementioned spectral clustering is concerned, scotch-taping is equivalent to adding weighted edges to the original graph.

\section{Formal solution}\label{sec:GreensFunction}
In this section, we derive a formal solution of Eq.~(\ref{EigenValueEquation}) using the Green's function formalism for eigenvector $\ket{\phi}_{k}$. 
Let us denote the $k$th generalized eigenvalue and eigenvector of the original graph as $\lambda^{0}_{k}$ and $\ket{\varphi}_{k}$, respectively, i.e., $\mat{L} \ket{\varphi}_{k} = (2-\lambda^{0}_{k}) \mat{D}^{0}_{U} \ket{\varphi}_{k}$, and we define the corresponding Green's function as 
\begin{align}
\mat{G}_{0k} \equiv \left( (2-\lambda^{0}_{k}) D^{0}_{U} - \mat{L}\right)^{-1}. \label{GF-1}
\end{align}
Then, Eq. (\ref{EigenValueEquation}) can be written as 
\begin{align}
& \left( \mat{G}^{-1}_{0k} - \widetilde{\mat{H}} \right) \ket{\phi}_{k} = \ket{0}, \notag\\
& \widetilde{\mat{H}} =  \lambda_{k} \mat{D}^{h}_{U} + \Delta\lambda_{k} \mat{D}^{0}_{U} - \sum_{r} \frac{2}{d^{v}_{r}} \ket{h}^{r}\ket{h}^{r \top}, \label{GF-2}
\end{align}
where $\Delta\lambda_{k} \equiv \lambda_{k} - \lambda^{0}_{k}$, and $\ket{0}$ is a vector in which all elements are equal to zero. 
If we analogously define the Green's function of the scotch-taped graph as $\mat{G}_{k} \equiv \left( \mat{G}^{-1}_{0k} - \widetilde{\mat{H}} \right)^{-1}$, we readily have the identity $\mat{G}_{k} = \mat{G}_{0k} + \mat{G}_{0k} \widetilde{\mat{H}} \mat{G}_{k}$ by definition. 
Then, Eq.~(\ref{GF-2}) yields 
\begin{align}
\ket{\phi}_{k} 
&= \ket{\varphi}_{k} + \mat{G}_{0k} \widetilde{\mat{H}} \ket{\phi}_{k} \notag\\
&= \ket{\varphi}_{k} + \mat{G}_{0k} \widetilde{\mat{H}} \ket{\varphi}_{k} + \mat{G}_{0k} \widetilde{\mat{H}} \mat{G}_{0k} \widetilde{\mat{H}} \ket{\varphi}_{k} + \cdots \notag\\
&= \ket{\varphi}_{k} + \mat{G}_{k} \widetilde{\mat{H}} \ket{\varphi}_{k}. \label{GF-3}
\end{align}
Here, the first term of the first equality is accounted for by the fact that $\ket{\varphi}_{k}$ is in the kernel of $G^{-1}_{0}$, and the second equality is obtained by recursively applying the first equality. 
We used the identity of $\mat{G}_{k}$ in the last equality. 
This is a variant of the Lippmann--Schwinger equation \cite{Ziman1969,CommentLippmannSchwinger}. 
The formal solution above shows how the low-dimensional representation, $\{ \varphi_{k} \}$, of the original graph is modified to $\{ \phi_{k} \}$ because of $\mat{H}$. 
In Appendix \ref{BrillouinWignerExpansion}, we show that a formal solution similar to Eq.~(\ref{GF-3}) can be obtained by the Brillouin--Wigner expansion \cite{Ziman1969}. 

The principle of scotch-taping is considerably simple, and the contribution of the external hyperedges is conceptually trivial. 
However, it is evident from this solution that, in general, the contribution of the external hyperedges can be highly complicated quantitatively.

\section{Solution of the stochastic block model}\label{sec:SBMSpectralMethod}
It is difficult to obtain further insight in a general setting. 
Therefore, we consider a random graph model called the stochastic block model (SBM) \cite{holland1983stochastic,WangWong87,Peixoto2017tutorial} and determine the conditions under which the signal of a module structure remains invariant, or becomes purely enhanced or weakened under the effect of the external hyperedges. 

The SBM is a random graph model with a planted (preassigned) module structure. 
In particular, we consider its microcanonical formulation \cite{Peixoto2017tutorial}. 
In this model, each node in a graph has a planted group assignment $\sigma \in \{1, \dots, K\}$, and the number of edges connecting nodes within/between groups $\sigma$ and $\sigma^{\prime}$ is specified as $e_{\sigma \sigma^{\prime}}$, which determines the strength of the module structure. 
The SBM generates a graph uniformly and randomly from instances that satisfy these constraints. 

We denote the physical nodes in group $\sigma$ as $U_{\sigma}$ ($|U_{\sigma}| = N_{\sigma}$) and the group label to which $i \in U$ belongs as $\sigma_{i}$. 
We also denote the factor nodes connecting physical nodes within/between groups $\sigma$ and $\sigma^{\prime}$ as $V^{0}_{\sigma \sigma^{\prime}}$, i.e., $\cup_{\sigma, \sigma^{\prime}} V^{0}_{\sigma \sigma^{\prime}} = V^{0}$ and $|V^{0}_{\sigma \sigma^{\prime}}| = e_{\sigma \sigma^{\prime}}$. 
Thus, the probability distribution of an incidence matrix $\mat{B}^{0}$ is expressed as 
\begin{align}
P(\mat{B}^{0}) 
= &\frac{1}{\mathcal{N}_{G}} 
\prod_{\sigma} \prod_{\alpha \in V_{\sigma \sigma}} \delta\left( \sum_{i \in U_{\sigma}}B^{0}_{i \alpha}, 2 \right) \notag\\
&\times \prod_{\sigma < \sigma^{\prime}} \prod_{\alpha \in V^{0}_{\sigma \sigma^{\prime}}} \delta\left( \sum_{i \in U_{\sigma}}B^{0}_{i \alpha}, 1 \right) \delta\left( \sum_{j \in U_{\sigma^{\prime}}}B^{0}_{j \alpha}, 1 \right), \label{microcanonicalSBM}
\end{align}
where $\delta(a,b)$ represents the Kronecker delta, and $\mathcal{N}_{G}$ is the total number of realizable graphs in the SBM; $\mathcal{N}_{G}$ is a normalization factor whose specific value need not be calculated for the present purposes. 
In the large graph limit, the degree of each node follows the Poisson distribution because the model only constrains the total number of edges within/between groups.

\subsection{Crude approximation and eigenvector invariance}\label{sec:CrudeApproximation}
Before attempting to obtain a precise solution of the SBM, we consider a crude approximation. 
Although this approximation does not allow us to investigate whether external hyperedges improve or deteriorate the resolution of module structures, it provides conditions under which the eigenvectors remain invariant under scotch-taping, i.e., $\mat{H}$ should not disturb the leading eigenvectors as noise.

Nodes in the same group are statistically equivalent in the SBM. 
Thus, as a crude approximation, we assume that an eigenvector element is well approximated by a group-wise constant, $\varphi_{ki} \approx \bar{\varphi}_{k \sigma}$, for any node $i \in U_{\sigma}$. 
In the absence of external hyperedges, eigenvalue equation (\ref{EigenValueEquation}) becomes
\begin{align}
\sum_{\sigma^{\prime}=1}^{K} f_{\sigma \sigma^{\prime}} \bar{\varphi}_{k \sigma^{\prime}} 
= (\lambda^{0}_{k}-1) \bar{\varphi}_{k \sigma}, 
\hspace{15pt}
f_{\sigma \sigma^{\prime}} \equiv \frac{1 + \delta(\sigma, \sigma^{\prime})}{c_{\sigma} N_{\sigma}} e_{\sigma \sigma^{\prime}}. \label{CrudeApproxUnperturbed}
\end{align}
We define the group-wise average degree as $c_{\sigma} \equiv \sum_{\sigma^{\prime}}  \left( 1 + \delta(\sigma, \sigma^{\prime}) \right) e_{\sigma \sigma^{\prime}}/N_{\sigma}$ (we denote the global average degree by $c \equiv 2 M_{0}/N$) and the $K \times K$ degree-corrected density matrix, $\mat{f}$, which is normalized as $\sum_{\sigma^{\prime}} f_{\sigma \sigma^{\prime}} = 1$. 
In the presence of external hyperedges, we have 
\begin{align}
&\sum_{\sigma^{\prime}} \left( c_{\sigma} f_{\sigma \sigma^{\prime}} 
+ 2\sum_{r} \frac{\overline{h^{r}_{\sigma}} \, \overline{h^{r}_{\sigma^{\prime}}} N_{\sigma^{\prime}} }{ \sum_{\sigma^{\prime\prime}} \overline{h^{r}_{\sigma^{\prime\prime}}} N_{\sigma^{\prime\prime}} }
\right) \bar{\phi}_{k \sigma^{\prime}} \notag\\
&\hspace{50pt}= \left( c_{\sigma}(\lambda_{k} - 1) + \lambda_{k} \sum_{r} \overline{h^{r}_{\sigma}} \right) \bar{\phi}_{k \sigma}, \label{CrudeApproxPerturbed}
\end{align}
where $\overline{h^{r}_{\sigma}} \equiv \sum_{i \in U_{\sigma}} h^{r}_{i}/N_{\sigma}$ ($0 \le \overline{h^{r}_{\sigma}} \le 1$). 
We approximated $\phi_{ki} \approx \bar{\phi}_{k \sigma}$ for any node $i \in U_{\sigma}$. 
Equations~(\ref{CrudeApproxUnperturbed}) and (\ref{CrudeApproxPerturbed}) are derived in Appendix \ref{sec:CrudeApproxDerivations}. 
In Eqs.~(\ref{CrudeApproxUnperturbed}) and (\ref{CrudeApproxPerturbed}), the trivial eigenvector with the largest eigenvalue, $\lambda^{0}_{1} = \lambda_{1} = 2$ \footnote{The Perron--Frobenius theorem ensures that this is the largest eigenvalue.}, is $(\bar{\varphi}_{11}, \dots, \bar{\varphi}_{1K}) = (\bar{\phi}_{11}, \dots, \bar{\phi}_{1K}) \propto \ket{1}_{K}$, where $\ket{1}_{K}$ is a $K$-dimensional column vector with all elements equal to unity. 
Therefore, all nontrivial eigenvectors are orthogonal to $\ket{1}_{K}$.

\begin{figure}[t!]
  \centering
  \includegraphics[width= 0.8\columnwidth, bb=0 0 350 280]{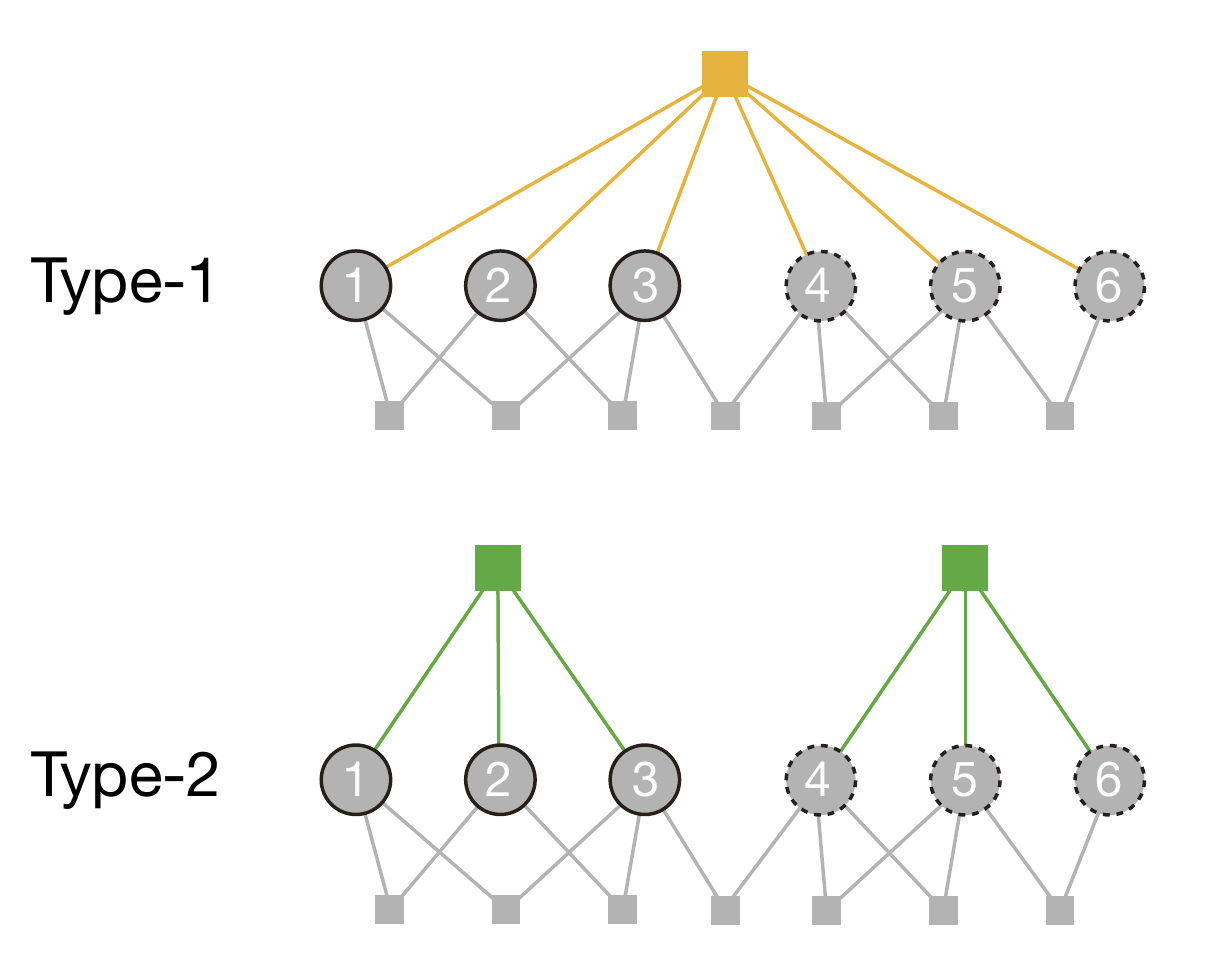}
  \caption{
  Schematic representations of Type-1 and Type-2 scotch-taping. 
  In both figures, circular nodes represent physical nodes, which are partitioned into two planted groups, indicated by the solid and dashed circles, respectively. 
  The bottom square nodes represent the factor nodes of the original graph, whereas the top square nodes represent the factor nodes corresponding to the annotations.
  }
  \label{fig:Type1Type2Schematic}
\end{figure}

We now consider two types of scotch-taping that leave an eigenvector in Eq.~(\ref{CrudeApproxPerturbed}) invariant. 
Examples are shown in Fig.~\ref{fig:Type1Type2Schematic}.

\begin{description}
\item[ Type-1 scotch-taping  ] 
The $r$th external hyperedge does not contribute to the left-hand side of Eq.~(\ref{CrudeApproxPerturbed}) if vector $(\overline{h^{r}_{1}} N_{1}, \dots, \overline{h^{r}_{K}} N_{K})$ is orthogonal to $(\bar{\phi}_{k1}, \dots, \bar{\phi}_{kK})$. 
In addition, when $(\sum_{r} \overline{h^{r}_{1}}, \dots, \sum_{r} \overline{h^{r}_{K}}) \propto (c_{1}, \dots, c_{K})$, the $k$th eigenvector remains invariant with respect to Eq.~(\ref{CrudeApproxUnperturbed}) although the eigenvalue is shifted. 
This is an interesting nontrivial case because an eigenvector is unaffected although the external hyperedge may be connected to physical nodes across the planted groups.  
This implies that a large-degree external hyperedge does not always adversely affect the structural information in the original graph. 
Hereafter, we refer to this case as Type-1 scotch-taping. 
\item[ Type-2 scotch-taping  ] 
Equation (\ref{CrudeApproxPerturbed}) also implies that, when $(\overline{h^{r}_{1}}, \dots, \overline{h^{r}_{K}})$ only has one nonzero element (i.e., it is one-hot shaped) for each hyperedge and $(\sum_{r} \overline{h^{r}_{1}}, \dots, \sum_{r} \overline{h^{r}_{K}}) \propto (c_{1}, \dots, c_{K})$, an eigenvector again remains invariant although the corresponding eigenvalue is shifted. 
Hereafter, we refer to this case as Type-2 scotch-taping. 
\end{description}

We note that the eigenvalues are shifted by these types of scotch-taping, even when the crude approximation is accurate. 
In the case of Type-1 scotch-taping, Eq.~(\ref{CrudeApproxPerturbed}) becomes 
\begin{align}
\sum_{\sigma^{\prime}} f_{\sigma \sigma^{\prime}} 
\bar{\phi}_{k \sigma^{\prime}}
= \left( \lambda_{k} - 1 + \lambda_{k} \kappa \right) \bar{\phi}_{k \sigma} 
\hspace{10pt} (k > 1), \label{CrudeApproxType1}
\end{align}
where $\kappa = \sum_{r} \overline{h^{r}_{\sigma}}/c_{\sigma}$, which is a constant irrespective of the group assignment $\sigma$ by the assumption of Type-1 scotch-taping. 
From a comparison between Eqs.~(\ref{CrudeApproxUnperturbed}) and (\ref{CrudeApproxType1}), for any $k>1$, 
\begin{align}
\lambda^{0}_{k}-1 &= \lambda_{k} - 1 + \lambda_{k} \kappa \notag\\
\lambda_{k} &= \frac{\lambda^{0}_{k}}{1 + \kappa}. \label{CrudeApproxType1-2}
\end{align}

In the case of Type-2 scotch-taping, Eq.~(\ref{CrudeApproxPerturbed}) becomes 
\begin{align}
\sum_{\sigma^{\prime}} f_{\sigma \sigma^{\prime}} 
\bar{\phi}_{k \sigma^{\prime}}
= \left( \lambda_{k} - 1 + \kappa (\lambda_{k} - 2) \right) \bar{\phi}_{k \sigma}. \label{CrudeApproxType2}
\end{align}
Then, similar to the Type-1 case, 
\begin{align}
\lambda^{0}_{k}-1 &= \lambda_{k} - 1 + \kappa (\lambda_{k} - 2) \notag\\
\lambda_{k} &= \frac{\lambda^{0}_{k}+2\kappa}{1 + \kappa} \label{CrudeApproxType2-2}
\end{align}
for any $k$. 
Although we are primarily interested in eigenvector invariance, the eigenvalues aid in evaluating the accuracy of the crude approximation. 

To confirm whether the present analysis provides an accurate estimate of the actual eigenvectors, let us consider a more specific parametrization of the SBM called the symmetric SBM \cite{AbbleReview2017}. 
This is an SBM of two equally sized planted groups with an assortative structure parametrized as $e_{11} = e_{22} =  e_{\mathrm{in}}$ and $e_{12} = e_{21} = e_{\mathrm{out}} \, (\le e_{\mathrm{in}})$. 
We use $\epsilon \equiv e_{\mathrm{out}}/(2e_{\mathrm{in}})$ to parametrize the strength of the module structure; a smaller value of $\epsilon$ indicates a stronger module structure. 
As an example of Type-1 scotch-taping, we consider an external hyperedge that is connected to all physical nodes (the top figure in Fig.~\ref{fig:Type1Type2Schematic}). 
As an example of Type-2 scotch-taping, we consider two external hyperedges: one connecting all the physical nodes belonging to group 1 and the other connecting all the physical nodes belonging to group 2 (the bottom figure in Fig.~\ref{fig:Type1Type2Schematic}). 

\begin{figure}[t!]
  \centering
  \includegraphics[width= 0.8\columnwidth, bb=0 0 464 988]{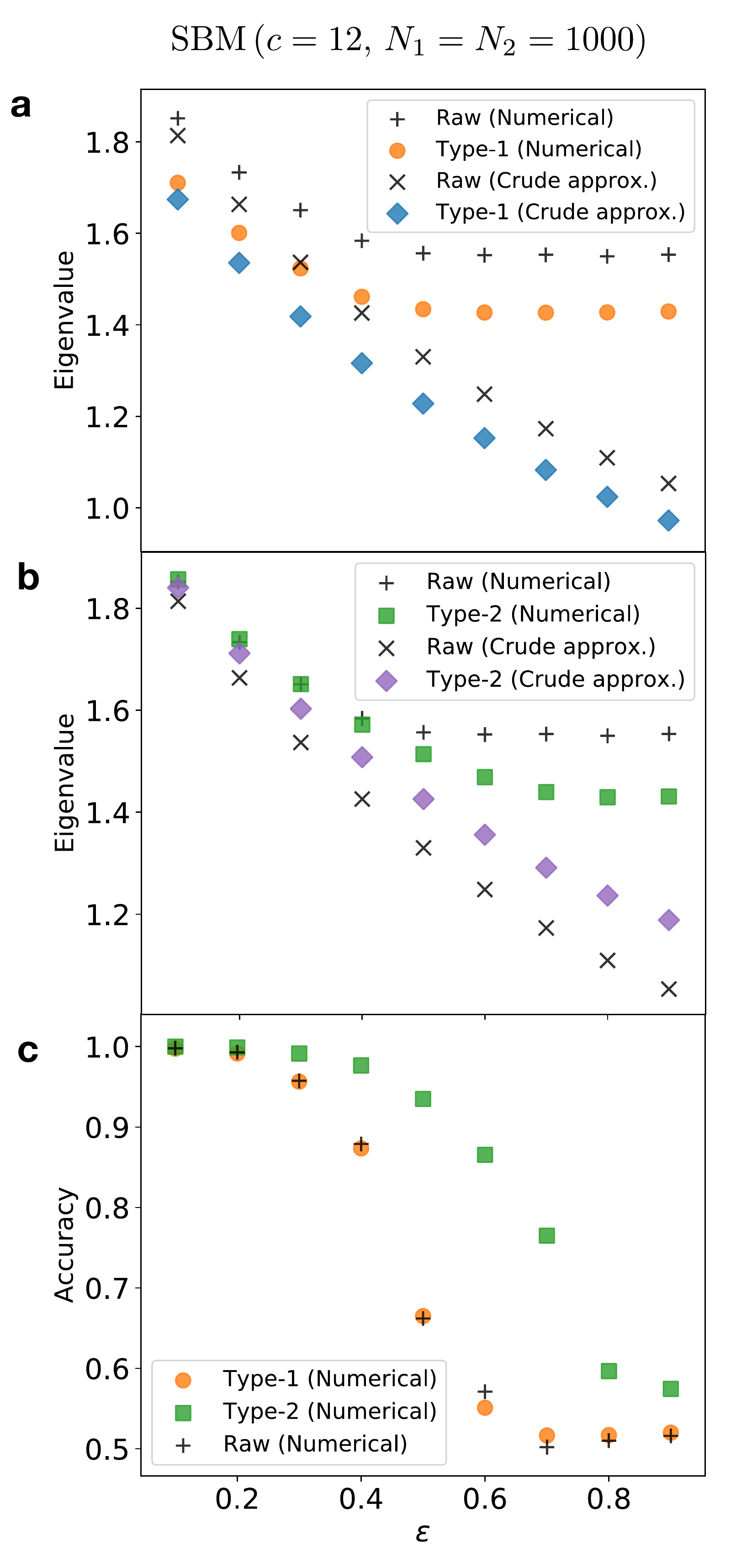}
  \caption{
  Second eigenvalues on a scotch-taped symmetric SBM ($c=12$, $N_{1}=N_{2}=1000$) with different strengths of the module structure:
  {\bf (a)} Type-1 and {\bf (b)} Type-2 scotch-taping. 
  {\bf (c)} Accuracy measured using the second eigenvectors for each scotch-taping. 
  In all the panels, the black crosses, yellow points, and green squares represent the eigenvalues obtained by numerical experiments on graphs without scotch-taping (Raw), Type-1 scotch-taped graphs (Type-1), and Type-2 scotch-taped graphs (Type-2), respectively. 
  Corresponding to these eigenvalues, the black tilted crosses represent $\lambda^{0}_{2}$ under the crude approximation, blue diamonds represent $\lambda_{2}$ in Eq.~(\ref{CrudeApproxType1-2}) ($\kappa=1/12$), and purple diamonds represent $\lambda_{2}$ in Eq.~(\ref{CrudeApproxType2-2}) ($\kappa=1/6$).
  }
  \label{fig:Eigenvalues}
\end{figure}

The second eigenvalues obtained in numerical experiments and the values predicted by the crude approximation are shown in Figs.~\ref{fig:Eigenvalues}{\bf a} and \ref{fig:Eigenvalues}{\bf b}. 
The eigenvalue estimate is not particularly accurate in general and becomes less accurate as $\epsilon$ increases. 
In particular, the crude approximation predicts that the second eigenvalue decreases monotonically; however, the actual second eigenvalue converges to a constant value. 

We then assess the second eigenvectors. 
Here, we characterize a second eigenvector by the accuracy of clustering, which is defined by the fraction of nodes for which the planted assignment that is correctly inferred from the signs of its elements; when the inference is completely random, the accuracy is $0.5$. 
In Fig.~\ref{fig:Eigenvalues}{\bf c}, we plot the obtained accuracy for the original, Type-1, and Type-2 scotch-taped graphs. 
Despite the low precision of the second eigenvalues, the accuracy for the original and that for the Type-1 scotch-taped graph are almost identical.  However, the accuracy for the original and that for the Type-2 scotch-taped graph are considerably different unless $\epsilon$ is very small. 

The inconsistency that was observed in the crude approximation after the application of Type-2 scotch-taping can be interpreted as follows. 
The eigenvector invariance under Type-2 scotch-taping indicates that although the scotch-taping enhances the signal of the planted group assignments, the second eigenvector is not improved further when it already exhibits a clear module structure (i.e., $\epsilon \approx 0$); otherwise, the approximation becomes invalid.

In the limit where $N \to \infty$, the non-leading eigenvalues constitute a spectral band; this is known as the ``semicircle law'' \cite{Mehta_RandomMatrices} and is due to the random nature of the graph. 
The second eigenvalue approaches the spectral band as $\epsilon$ increases, and when the second eigenvalue is no longer isolated from the spectral band, the graph becomes indistinguishable from a uniform random graph in terms of its spectrum. 
This phenomenon is known as detectability phase transition \cite{NadakuditiNewman2012,Kawamoto2015Laplacian,AbbleReview2017,MooreReview2017}. 
This, in fact, accounts for the convergence of the second eigenvalue in Figs.~\ref{fig:Eigenvalues}{\bf a} and \ref{fig:Eigenvalues}{\bf b}; the plateau indicates that the second eigenvalue reached the edge of the spectral band. 
Scotch-taping acts as noise if it promotes the occurrence of detectability phase transition. 
Unfortunately, as we confirmed in Fig.~\ref{fig:Eigenvalues}, we cannot derive the spectral band from the crude approximation. 

A flaw of the crude approximation is that the fluctuation of the eigenvector elements is neglected. 
This fluctuation is essential for detectability phase transition and should also be related to the inconsistency that we observed after applying Type-2 scotch-taping because the fluctuation effect becomes prominent when the module structure is weak \cite{Kawamoto2015Laplacian}.

\subsection{Message-passing equation}\label{sec:MessagePassingEqn}
To account for the fluctuation of eigenvector elements, we solve the corresponding equation averaged over Eq.~(\ref{microcanonicalSBM}). 
Hereafter, we focus on the second eigenvalue and eigenvector in the large graph limit ($N \gg 1$). 
We also consider the situation where the number of external hyperedges $R$ is $o(N)$, although each external hyperedge can be connected to $O(N)$ physical nodes; if $R$ were as large as $N$, the contribution of the external hyperedges would trivially be dominant, and thus we could no longer regard the original graph as a primary object. 

We begin with the following formulation for the second eigenvalue, $\lambda_{2} = 2 s^{2}_{2}$:
\begin{align}
&\lambda_{2} = 
\max_{\ket{z}} \frac{2}{N} \bra{z} \mat{\mathsf{B}} \mat{\mathsf{B}}^{\top} \ket{z}, \notag\\
&\hspace{10pt} \text{subject to} \hspace{10pt} 
\bra{z}\ket{z} = N \hspace{10pt} \text{and} \hspace{10pt}
\bra{1}_{N}\mat{D}^{1/2}_{U}\ket{z} = 0. \label{lambda2Maximization1}
\end{align}
The second constraint represents the orthogonality condition relative to the first eigenvector. 
As considered in Sec.~\ref{sec:SpectralFramework}, we transform the variable as $\ket{x} = \mat{D}^{-1/2}_{U} \ket{z}$ and rewrite the maximization function using the internal structure of $\mat{\mathsf{B}}$. 
Then, Eq.~(\ref{lambda2Maximization1}) is reformulated as 
\begin{align}
& \lambda_{2} = \lim_{\beta\to\infty} \frac{2}{\beta N} 
 \mathop{\mathrm{extr}}_{\lambda, \gamma} \left\{ \log Z(\beta, \lambda, \gamma)\right\}, \label{SecondEigenvalue}\\
& Z(\beta, \lambda, \gamma) = \int d\ket{x} \, \mathrm{e}^{\beta E(\ket{x}, \lambda, \gamma)}, \label{PartitionFunction}\\
& E(\ket{x}, \lambda, \gamma) 
= \bra{x} \mat{B} \mat{D}^{-1}_{V} \mat{B}^{\top} \ket{x} \notag\\
&\hspace{50pt}-\frac{\lambda}{2} \left( \bra{x}\mat{D}_{U}\ket{x} - N \right) - \gamma \bra{1}_{N}\mat{D}_{U}\ket{x}, \label{EnergyFunction}
\end{align}
where $\mathrm{extr}$ represents the extremization, and $\lambda$ and $\gamma$ are Lagrange multipliers. These multipliers should be $\gamma=0$ and $\lambda = \lambda_{2}$ such that the saddle-point condition for $E(\ket{x}, \lambda, \gamma)$ yields the eigenvalue equation. 
The second eigenvector, $\ket{\phi}_{2}$, appears as the saddle point with respect to $\ket{x}$ in Eq.~(\ref{SecondEigenvalue}). 

We are interested in the configuration average over the realizations of matrix $\mat{B}^{0}$, which is specified by Eq.~(\ref{microcanonicalSBM}), and we denote this average by $[\cdots]_{\mat{B}^{0}}$. 
Thus, our goal is to determine $\left[ \lambda_{2} \right]_{\mat{B}^{0}}$. 
Here, assuming that the configuration average can be interchanged  with the limit with respect to $\beta$ and the extremization of $\lambda$ and $\gamma$, the replica trick yields the following expression for $\left[ \lambda_{2} \right]_{\mat{B}^{0}}$: 
\begin{align}
\left[ \lambda_{2} \right]_{\mat{B}^{0}} 
= \lim_{\beta\to\infty} \lim_{n \to 0} \frac{1}{\beta N} \mathop{\mathrm{extr}}_{\lambda, \gamma} \frac{\partial}{\partial n} \log \left[ Z^{n}(\beta, \mu, \lambda, \gamma) \right]_{\mat{B}^{0}}. \label{ReplicaEqn}
\end{align}
The detailed calculation of Eq.~(\ref{ReplicaEqn}) is presented in Appendix \ref{sec:ReplicaMethod}. 

From the saddle-point estimate in Eq.~(\ref{ReplicaEqn}), we obtain a self-consistent equation of the eigenvector elements. 
For the elements corresponding to the physical nodes in group $\sigma$, we denote the distribution of the eigenvector elements as $Q_{\sigma}\left( {\sf x} \right)$, i.e., 
\begin{align}
Q_{\sigma}\left( {\sf x} \right) = \frac{1}{N_{\sigma}} \sum_{i \in U_{\sigma}} \delta\left( {\sf x} - \phi_{2i} \right), 
\end{align}
and we parametrize it using a Gaussian mixture as follows: 
\begin{align}
Q_{\sigma}\left( {\sf x} \right) 
= \int d\mathsf{A} d\mathsf{H} \, q_{\sigma}(\mathsf{A}, \mathsf{H}) \, 
\sqrt{\frac{\beta \mathsf{A}}{2\pi}} \mathrm{e}^{-\frac{\beta \mathsf{A}}{2} (\mathsf{x} - \mathsf{H})^{2}}. 
\end{align}
Here, $q_{\sigma}(\mathsf{A}, \mathsf{H})$ is the mixture weight of the Gaussian distribution with mean $\mathsf{H}$ and precision parameter $\mathsf{A}$. 
The saddle-point estimate in Eq.~(\ref{ReplicaEqn}) yields the following self-consistent (message passing) equation with respect to $q_{\sigma}(\mathsf{A}, \mathsf{H})$: 
\begin{widetext}
\begin{align}
&q_{\sigma}\left( \mathsf{A}, \mathsf{H} \right) 
= \sum_{\ket{h}} \mathsf{P}_{\sigma}(\ket{h}) 
\sum_{d=0}^{\infty} \mathcal{P}_{c_{\sigma}}\left(d\right) 
\prod_{\ell=1}^{d}\left( \sum_{\sigma^{\prime}} f_{\sigma \sigma^{\prime}} 
\int d\mathsf{A}_{\ell}d\mathsf{H}_{\ell} \, q_{\sigma^{\prime}}\left( \mathsf{A}_{\ell}, \mathsf{H}_{\ell} \right) \right)
\notag\\
&\times \delta\left( \mathsf{A} - \lambda (d + \sum_{r} h^{r}) + \sum_{\ell=1}^{d} \left( 1 - \frac{1}{\lambda + \mathsf{A}_{\ell}} \right)^{-1} \right) 
\delta\left( \mathsf{H} - \frac{ 2 \sum_{r} \frac{N}{d^{v}_{r}} h^{r}m_{r} - \sum_{\ell=1}^{d} \frac{ \mathsf{A}_{\ell}\mathsf{H}_{\ell}}{1-(\lambda + \mathsf{A}_{\ell})} }{\lambda (d + \sum_{r} h^{r}) - \sum_{\ell=1}^{d} \left( 1 - \frac{1}{\lambda + \mathsf{A}_{\ell}} \right)^{-1}} \right). \label{PhysicalCavityEqMain}
\end{align}
\end{widetext}
Here, $\delta(\cdot)$ represents the Dirac delta. 
Further, $\mathcal{P}_{c_{\sigma}}\left(d\right)$ is a Poisson distribution with mean $c_{\sigma}$, which represents the degree distribution of a physical node in group $\sigma$. 
$\mathsf{P}_{\sigma}(\ket{h})$ is the empirical distribution of the external hyperedges, which is defined as 
\begin{align}
\mathsf{P}_{\sigma}(\ket{h}) 
&\equiv \frac{1}{N_{\sigma}} \sum_{i \in U_{\sigma}} 
\prod_{r} \delta\left(h^{r}, h^{r}_{i} \right), \label{EmpiricalDistribution}
\end{align}
where $\ket{h} = (h^{1}, \dots, h^{R})$ is an indicator vector; the product in Eq.~(\ref{EmpiricalDistribution}) evaluates the set of factor nodes to which a physical node is connected. 
Moreover, in Eq.~(\ref{PhysicalCavityEqMain}), 
\begin{align}
m_{r} \equiv \frac{1}{N} \sum_{i \in U} h^{r}_{i} \phi_{2i} 
\end{align}
is the inner product of the $r$th external hyperedge and the second eigenvector. 
In physics terminology, $m_{r}$ is an external magnetic field. 
The eigenvector-element distribution is essentially characterized by the distribution of $\mathsf{H}$ in $q_{\sigma}\left( \mathsf{A}, \mathsf{H} \right)$ because $\beta \to \infty$; however, $\mathsf{A}$ should not be neglected because it affects $\mathsf{H}$. 
In fact, in the absence of external hyperedges, Eq.~(\ref{CrudeApproxUnperturbed}) can be derived from Eq.~(\ref{PhysicalCavityEqMain}) in the limit where $\mathsf{A} \gg 1$ (see Appendix \ref{sec:SmallFluctuationLimit}). 

Equation (\ref{PhysicalCavityEqMain}) is a self-consistent equation that fully considers the structure and statistics of the SBM, which are reflected by $f_{\sigma \sigma^{\prime}}$ and $\mathcal{P}_{c_{\sigma}}\left(d\right)$, respectively, as well as the distribution of external hyperedges $\mathsf{P}_{\sigma}(\ket{h})$. 
The saddle-point conditions in Eq.~(\ref{ReplicaEqn}) also yield a self-consistent equation for $m_{r}$ as well as an equation for $\lambda$ (see Appendix \ref{sec:ReplicaMethod}). 
Two corrections to the eigenvector-element distribution are $2N \sum_{r} h^{r}m_{r}/d^{v}_{r}$ and $d + \sum_{r} h^{r}$, corresponding to the correction terms (owing to the external hyperedges) on the left- and the right-hand sides of Eq.~(\ref{CrudeApproxPerturbed}). 
The case of $m_{r}=0$ corresponds to the orthogonality condition in Type-1 scotch-taping. 


We recall that Eq.~(\ref{GF-3}) is a formal solution for a general scotch-taped graph, requiring recursive operations of matrix inverse $G_{0k}$ and $\widetilde{\mat{H}}$, which includes $\lambda_{k}$ and $\{\ket{h}^{r}\}$. 
Although Eq.~(\ref{PhysicalCavityEqMain}) is somewhat analogous to Eq.~(\ref{GF-3}), we observe in the next section that it has better interpretability. 

\section{Analysis of scotch-taping using the message-passing equation}\label{AnalysisMessagePassing}
Equation (\ref{PhysicalCavityEqMain}) is substantially more informative than the crude approximation. 
For example, although Type-1 scotch-taping is apparently harmless, Eq.~(\ref{PhysicalCavityEqMain}) indicates that a structural signal by the eigenvector would eventually be weakened if we attached sufficiently many external hyperedges of Type 1 (Sec.~\ref{sec:Type1Analysis}). 
It also explains how Type-2 scotch-taping improves the resolution of module structure when $\epsilon$ is not small (Sec.~\ref{sec:Type2Analysis}). 
In addition, although one may speculate that splitting an external hyperedge with a large degree into a set of several hyperedges with smaller degree may be an effective strategy, Eq.~(\ref{PhysicalCavityEqMain}) indicates that, generally, neither strategy is superior (Sec.~\ref{sec:R-degreeDependence}).

\subsection{Uniform external hyperedges}\label{sec:Type1Analysis}
We first analyze the contribution of external hyperedges such that each factor node is connected to all physical nodes. 
As discussed in Sec.~\ref{sec:CrudeApproximation}, we consider the symmetric SBM with an assortative structure ($0 < \epsilon \le 1$). 
Then, this is Type-1 scotch-taping. 
Each external hyperedge satisfies the condition for orthogonality to the nontrivial leading eigenvectors in the crude approximation and $m_{r}=0$ in the message-passing equations. 

\begin{figure}[t!]
  \centering
  \includegraphics[width= 0.8\columnwidth, bb=0 0 546 665]{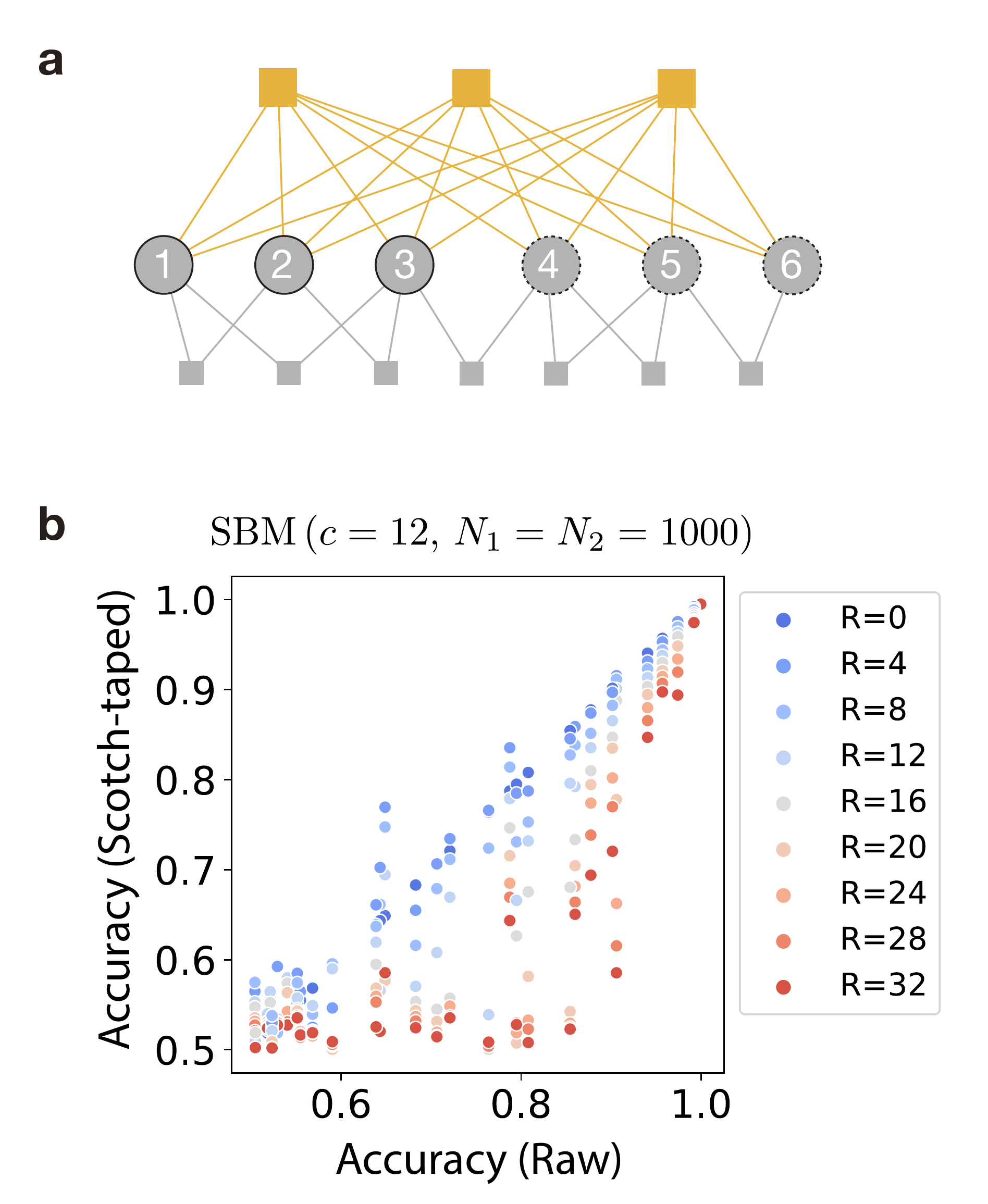}
  \caption{
  ({\bf a}) Schematic of multiple external hyperedges of Type 1, and ({\bf b}) a scatter plot that compares the accuracy values of the spectral clustering for the original graphs (Raw) and their scotch-taped counterparts (Scotch-taped). 
  The original graphs are generated from the symmetric SBM ($c=12$, $N_{1}=N_{2}=1000$). 
  The plot is obtained through numerical experiments with different numbers of external hyperedges $R$ and various values of $\epsilon$, ranging from $\epsilon = 0.1$ to $0.9$. 
  }
  \label{fig:OverlapComparisons}
\end{figure}

The crude approximation implies that the second eigenvector remains invariant under Type-1 scotch-taping. 
When we have only one external hyperedge, we have confirmed that this is apparently correct for the symmetric SBM (Fig.~\ref{fig:Eigenvalues}). 
However, this invariance should be violated when the number of hyperedges $R$ is sufficiently large. 
The results of a numerical experiment, as shown in Fig.~\ref{fig:OverlapComparisons}, demonstrate that the spectral clustering of the scotch-taped graph is less correlated with the planted group assignments than that of the original graph for large $R$.

\begin{figure}[t!]
  \centering
  \includegraphics[width= \columnwidth, bb = 0 0 1307 1218]{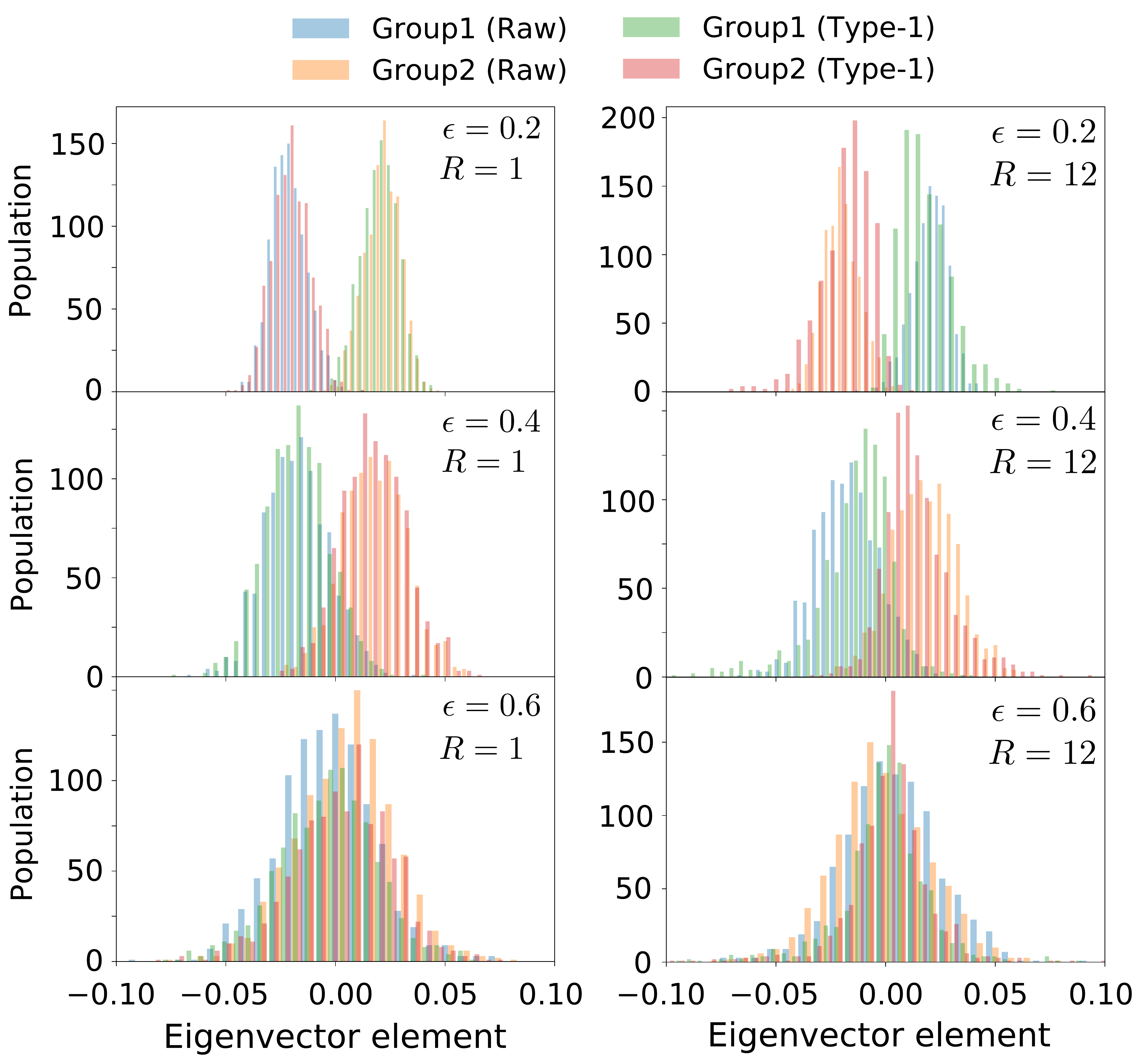}
  \caption{
  Populations of the second eigenvector elements of the symmetric SBM instances (Raw) and scotch-taped instances of Type 1 (Type-1). 
  We use the same graph instances as in Fig.~\ref{fig:Eigenvalues}. 
  The strength of the module structure ($\epsilon$) and the number of external hyperedges ($R$) are shown in each panel. 
  The populations of elements corresponding to groups 1 and 2 in the original graphs are represented in blue and yellow, respectively. 
  The populations of elements corresponding to groups 1 and 2 in the scotch-taped graphs are represented in green and red, respectively. 
  }
  \label{fig:Assortative_Type1}
\end{figure}

To better observe this phenomenon, the distributions of the second eigenvector elements obtained using numerical experiments are plotted in Fig.~\ref{fig:Assortative_Type1}. 
When the second eigenvector is significantly correlated with the planted group assignments, the distribution is bimodal; the elements corresponding to the physical nodes in group 1 constitute one peak, and those corresponding to the physical nodes in group 2 constitute the other peak. 
When the distribution is unimodal, the spectral clustering can no longer distinguish the scotch-taped graph from a uniform random graph. 

We analyze the behavior observed in Figs.~\ref{fig:OverlapComparisons} and \ref{fig:Assortative_Type1} using the message-passing equation (\ref{PhysicalCavityEqMain}). 
We note that $h^{r}=1$ for all $r$ because a physical node is always incident to the $r$th external hyperedge. 
To simplify this argument, we use a regular approximation and replace the degree $d$ by the average degree $c$ of the original graph. 
We also assume that $A_{Q}$ is a constant denoted by $a$. 
This is known as effective medium approximation \cite{Kawamoto2015Laplacian}. 
Then, the updating part with respect to $A_{Q}$ (i.e., the constraint of the former delta function) in Eq.~(\ref{PhysicalCavityEqMain}) yields 
\begin{align}
& a + \frac{c_{\sigma}}{\lambda - 1 + a} = c_{\sigma} (\lambda - 1) + \lambda R. \label{Type1AQ}
\end{align}
Although $\lambda$ can vary with $a$, $\lambda$ is bounded because $\lambda$ corresponds to the eigenvalue $\lambda_{2}$. 
Moreover, owing to the spectral band, $\lambda_{2}$ cannot be excessively small. 
Therefore, the right-hand side of Eq.~(\ref{Type1AQ}) is dominated by the second term when $R$ is sufficiently larger than $c_{\sigma}$. 
Consequently, $a$ should monotonically increase as $R$ increases. 

We now consider the updating part with respect to $H_{Q}$ (i.e., the constraint of the latter delta function) in Eq.~(\ref{PhysicalCavityEqMain}). 
Because $\gamma = 0$ and $m_{r} = 0$, the delta function is reduced to 
\begin{align}
\delta\left( H_{Q} - \frac{1}{\lambda - 1 + a} \sum_{\ell=1}^{c_{\sigma}} H_{Q\ell} \right). \label{Type1UpdateEq}
\end{align}
Here, $a$ plays the role of a global shrinkage parameter. 
When $a$ is not excessively large, the distribution $q_{\sigma}\left( H_{Q} \right)$ has a peak at a nonzero value of $H_{Q}$ as the fixed point of the message-passing equation ($H_{Q \ell}$ in Eq.~(\ref{Type1UpdateEq}) is sampled from both groups 1 and 2 based on $f_{\sigma \sigma^{\prime}}$ and $q_{\sigma^{\prime}}\left( H_{Q \ell} \right)$). 
However, when $a$ is sufficiently large, the distribution $q_{\sigma}\left( H_{Q} \right)$ that peaks at the origin is the only solution. 

In summary, it was demonstrated that uniform external hyperedges promote the occurrence of detectability phase transition (i.e., deteriorate the resolution of spectral clustering) when $R$ is sufficiently larger than the average degree. 
This behavior can indeed be confirmed in Fig.~\ref{fig:OverlapComparisons} (bottom).

\subsection{External hyperedges consistent with the planted module structure}\label{sec:Type2Analysis}
Let us consider external hyperedges such that each external factor node is connected to a set of physical nodes sharing a group assignment. 
We again consider the symmetric SBM with an assortative structure. 
Then, this is Type-2 scotch-taping. 
Here, we let $R_{\sigma}$ be the number of external hyperedges connected to the physical nodes in group $\sigma$, and denote their labels as $r = (\sigma, r^{\prime})$ ($r^{\prime} \in \{1, \dots R_{\sigma}\}$ and $\sum_{\sigma} R_{\sigma} = R$). 

\begin{figure}[t!]
  \centering
  \includegraphics[width= \columnwidth, bb = 0 0 1331 1242]{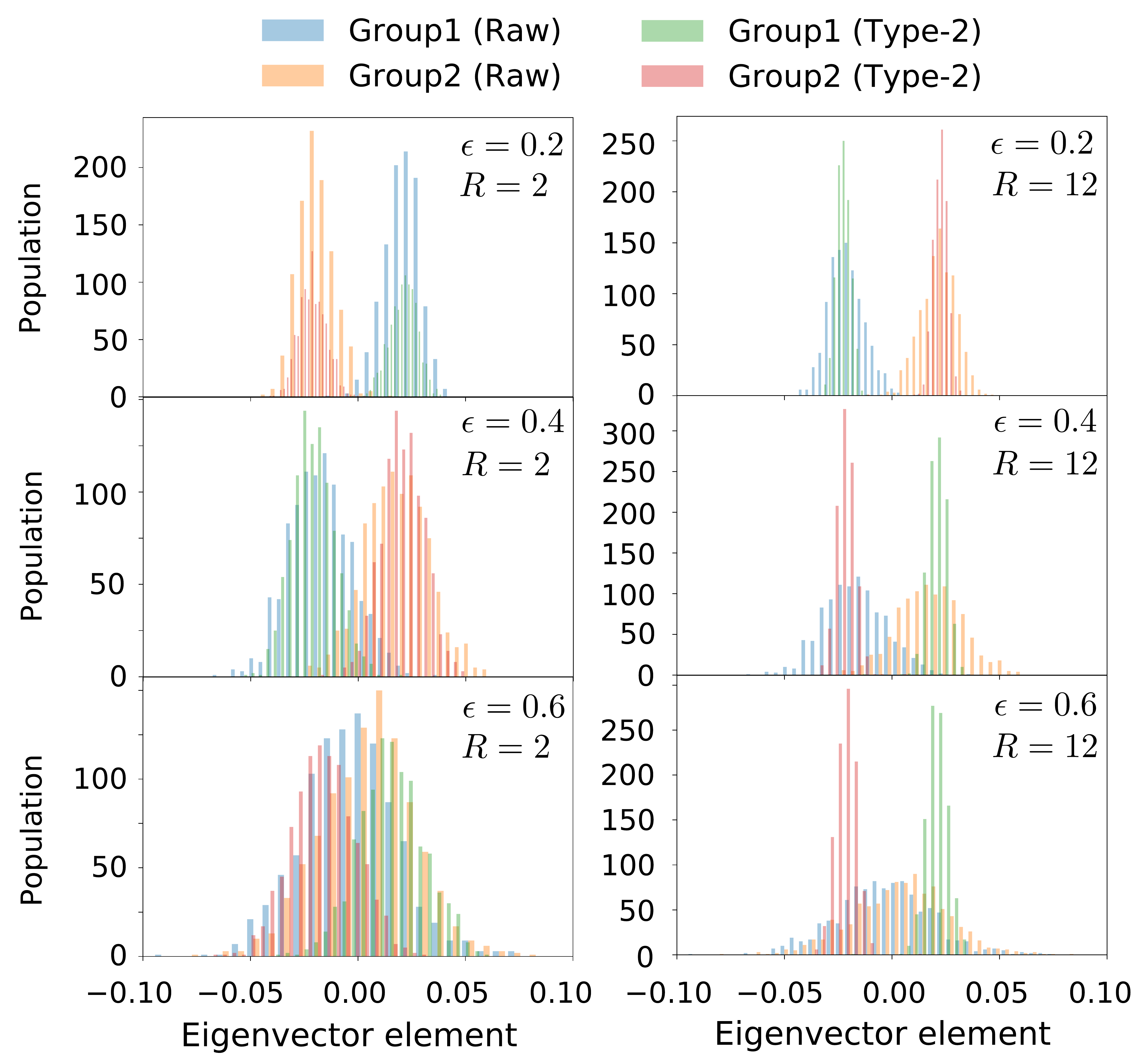}
  \caption{
  Populations of the second eigenvector elements of the symmetric SBM instances (Raw) and scotch-taped instances of Type 2 (Type-2). 
  We use the same graph instances as in Fig.~\ref{fig:Eigenvalues}. 
  The plots are drawn in the same manner as in Fig.~\ref{fig:Assortative_Type1}. 
  }
  \label{fig:Assortative_Type2}
\end{figure}

The distributions of the second eigenvector element for Type-2 scotch-taping are shown in Fig.~\ref{fig:Assortative_Type2}. 
By applying the same approximation as in the previous section, we obtain the same argument as in Eq.~(\ref{Type1AQ}) for the behavior of the parameter $a$. 
We note that $d^{v}_{(\sigma, r^{\prime})} = N_{\sigma}$, and $\mathsf{P}_{\sigma}(\ket{h})$ is a pair of delta functions that have peaks at $\ket{h}$ such that $h^{r} = 1$ for $r \in \{ (\sigma, r^{\prime}) \}$ and $h^{r} = 0$ otherwise (for $\sigma=1$ and $\sigma=2$, respectively). 
Thus, for the updating part with respect to $H_{Q}$ in Eq.~(\ref{PhysicalCavityEqMain}), 
\begin{align}
\delta\left( H_{Q} - \frac{1}{a} \left( 2 \sum_{r \in \{ (\sigma, r^{\prime}) \}} \frac{N}{N_{\sigma}} m_{r} + \frac{a}{\lambda - 1 + a} \sum_{\ell=1}^{d} H_{Q\ell} \right) \right). \label{Type2UpdateEq}
\end{align}
Although the increase in the parameter $a$ again reduces the overall scale, as $H_{Q}$ is ``pinned'' by the terms with $m_{r}$, the eigenvector elements remain polarized. 
As the number of external hyperedges increases, the contribution from $\sum_{\ell=1}^{d} H_{Q\ell}$ becomes negligible (because each term in the sum becomes small owing to the increase in $a$), and the distribution $q_{\sigma}\left( H_{Q} \right)$ is dominated by the terms with $m_{r}$. 
Consequently, the fluctuation of $H_{Q}$ is suppressed. 
The pinning effect and the variance reduction for $q_{\sigma}\left( H_{Q} \right)$ explain the resolution improvement in spectral clustering. 
Such a behavior is indeed confirmed in Fig.~\ref{fig:Assortative_Type2}. 

\subsection{Few external hyperedges with large degrees vs. many external hyperedges with small degrees}\label{sec:R-degreeDependence}

\begin{figure*}[ht!]
  \centering
  \includegraphics[width= 2\columnwidth, bb = 0 0 2011 690]{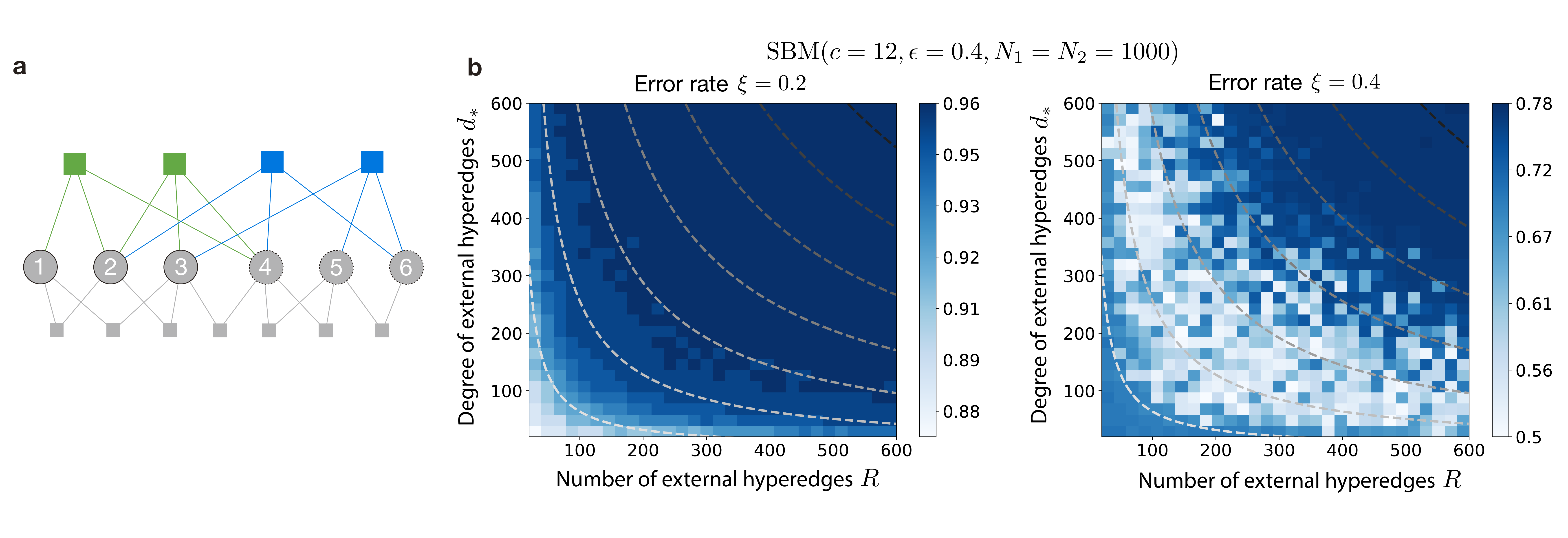}
  \caption{Numerical experiments regarding spectral clustering on the symmetric SBM with noisy external hyperedges. 
  ({\bf a}) A schematic of the scotch-taping we consider, and ({\bf b}) the density plots of the accuracy values for various numbers of external hyperedges $R$ (horizontal axis) and the degree of each external hyperedge $d_{\ast}$ (vertical axis). 
  We used the same graph instance of the SBM as in Fig.~\ref{fig:Eigenvalues}. 
  The error rate of the noisy external hyperedges is represented by $\xi$.
  The dashed curves indicate the parameter pairs for which the total number (represented by the color depth) of edges among the external hyperedges is preserved. 
  }
  \label{fig:OverlapDensityPlot}
\end{figure*}

We assume that an annotation label is shared by a large number of physical nodes. 
In previous sections, we considered scotch-taping such that each factor node is connected to all these physical nodes. 
However, a division into several external hyperedges with small degree may be more beneficial. 
We consider this problem in this section. 

In fact, neither of these strategies has a general benefit or drawback. 
As mentioned above, the terms relevant to the external hyperedges are $2 N \sum_{r} h^{r} m_{r}/d^{v}_{r}$ and $\sum_{r=1}^{R} h^{r}$. 
The term $2 N \sum_{r} h^{r} m_{r}/d^{v}_{r}$ in the message-passing equation (\ref{PhysicalCavityEqMain}) is always $O(1)$ or less; in order that $h^{r} m_{r}$ be $O(1)$, the degree $d^{v}_{r}$ should be $O(N)$, whereas $h^{r} m_{r}$ can be at most $O(1/N)$ when $d^{v}_{r}$ is $O(1)$. 
Thus, dividing an external hyperedge into multiple hyperedges with small degree does not necessarily strengthen or weaken its contribution. 

The contribution from $\sum_{r=1}^{R} h^{r}$ is also preserved as long as the total number of external edges remains the same. 
For example, let us assume that we originally have two external hyperedges corresponding to groups 1 and 2, and node $i \in U$ is incident to the external hyperedge corresponding to group 1; then, we would have $(h^{1}_{i}, h^{2}_{i}) = (1, 0)$, which is the $i$th row in the $H$ matrix in Eq.~(\ref{TotalIncidenceMatrix}). 
If we divide the external hyperedge corresponding to group 1 into two, then we would have $(h^{(1,1)}_{i}, h^{(1,2)}_{i}, h^{2}_{i})$ that is equal to either $(1,0,0)$ or $(0,1,0)$. 
In all cases, we have $\sum_{r=1}^{R} h^{r}_{i} = 1$. 
This demonstrates that the statistic of $\sum_{r=1}^{R} h^{r}$ is invariant under the divisions of external hyperedges. 

We further confirm that there is no clear benefit between few external hyperedges with large degrees and many external hyperedges with small degrees through a numerical experiment. 
In this experiment, we consider noisy external hyperedges of Type 2 with  error rate $\xi$, as schematically shown in Fig.~\ref{fig:OverlapDensityPlot}{\bf a}. 
That is, we add $R/2$ external hyperedges to increase the density in group 1, and $R/2$ external hyperedges to increase the density in group 2. 
Among $d^{v}_{r} = d_{\ast} \, (\mathrm{const.})$ edges incident to an external hyperedge of the first half, edges are randomly connected to the physical nodes in group 1 with probability $1-\xi$, and the rest are randomly connected to the physical nodes in group 2. 
Similarly, for an external hyperedge of the second half, edges are randomly connected to the physical nodes in group 2 with probability $1-\xi$, and the rest are randomly connected to the physical nodes in group 1. 

Each panel in Fig.~\ref{fig:OverlapDensityPlot}{\bf b} shows the accuracy regarding the planted group assignments of the symmetric SBM as the number of external hyperedges (horizontal axis) and the degree of each external hyperedge (vertical axis) vary. 
The dashed curves represent the parameter pairs for which $d_{\ast} R$ is conserved. 
Evidently, the accuracy is the same on each dashed curve, implying that there is no preference in the balance between the number of external hyperedges $R$ and the degree of each external hyperedge $d_{\ast}$. 
Interestingly, it is also confirmed from the right panel of Fig.~\ref{fig:OverlapDensityPlot}{\bf b} that the accuracy improvement by scotch-taping can be nonmonotonic. 
Therefore, depending on the values of $\epsilon$ and $\xi$, there exists an intermediate range in the $(d_{\ast}, R)$ space where the external hyperedges act as noise even though this scotch-taping eventually improves the accuracy when we add more external hyperedges.

\section{Discussion}\label{sec:Discussion}
We considered a simple method to encode node annotations on a graph as a factor graph, and we established the mathematical basis of the method in the spectral framework. 
Even though scotch-taping may be used in various inference problems, we focused on the inference of an assortative module structure. 

As mentioned in Sec.~\ref{sec:FactorGraph}, because scotch-taping is based only on the data representation, it can be combined with an arbitrary algorithm on graphs. 
Non-negative matrix factorization (NMF) \cite{Lee1999_NMF} is a method similar to spectral clustering, and in Appendix \ref{sec:NMF}, we present applications of NMF to the scotch-taped incidence matrix $\mat{B}$. 
A numerical experiment in Appendix \ref{sec:NMF} demonstrates that, even a few uniform external hyperedges disrupt the inference of the module structure of the original graph, indicating that the observed behavior regarding spectral clustering in Secs.~\ref{sec:CrudeApproximation} and \ref{sec:Type1Analysis} is not universal.  

Scotch-taped graphs can be used as input to nonlinear graph embedding methods, such as DeepWalk \cite{Perozzi2014DeepWalk}, node2vec \cite{grover2016node2vec}, and LINE \cite{Tang2015LINE}. 
The fact that these methods can only take a graph as input is occasionally characterized as their fundamental limitation \cite{hamilton2017representation}. 
However, scotch-taping naturally extends the applicability of these embedding methods. 


The graph convolutional network (GCN) \cite{KipfWelling2016} is a popular GNN algorithm. 
Although the GCN already considers node annotations (or features) as well as the graph structure, it is also possible to use a scotch-taped graph as input. 
The original GCN uses (monopartite) graphs; however, we can immediately generalize it to factor graphs (bipartite graphs) by considering the following variant of the feed-forward architecture of the GCN: 
\begin{align}
\mat{X}^{t+1} = \hat{\sigma} \left( \mat{\mathsf{B}} \, \hat{\sigma}\left( \mat{\mathsf{B}}^{\top} \mat{X}^{t} \mat{W}_{U}^{t} \right) \mat{W}_{V}^{t} \right), \label{bipartiteGCN}
\end{align}
where $t \in \{0, 1, \dots, T\}$ represents the layer index, and $\hat{\sigma}$ is a nonlinear operator. 
$\mat{X}^{t} \in \mathbb{R}^{N \times R^{t}_{U}}$ is the feature matrix of the physical nodes ($\mat{X}^{0} = \mat{H}$), which is to be updated across the layers. 
$\mat{W}_{U}^{t}$ and $\mat{W}_{V}^{t}$ are linear transforms at the $t$th layer that are to be learned. 
Equation (\ref{bipartiteGCN}) has the potential to encode richer information because two different types of node attributes can be inserted into $\mat{X}^{0}$ and $\mat{\mathsf{B}}$, respectively. 
We empirically confirmed that scotch-taping can both improve and deteriorate prediction in nonlinear graph embedding methods and GCN, depending on the dataset.

Let us finally consider some qualitative distinctions between the inference using a scotch-taped graph and other popular methods. 
In Bayesian inference \cite{NewmanClauset2016,Hric_PRX2016}, graph data and node annotations interact indirectly in a generative model, because the former contributes to a likelihood whereas the latter does to a prior. 
In contrast, scotch-taping treats both graph data and node annotations on an equal footing; thus, they interact more directly.
Similarly, in the GCN framework, the contribution of node attributes in the feature matrix in $\mat{X}^{0}$ is different from that in the incidence matrix $\mat{\mathsf{B}}$.

What scotch-taping suggests is \textit{why don't you simply add edges if you believe a set of nodes are similar or dissimilar to each other?}. This is simplistic and may even appear ad hoc. 
However, it is certainly a choice when other sophisticated methods fail. 
In real data analysis, it is conceivable that several practitioners have used a technique similar to scotch-taping. 
In any case, we can always consider using scotch-taping to further improve the performance of an algorithm, or to determine whether node annotations are consistent with the underlying graph structure.

\begin{acknowledgments}
This study was partly supported by the New Energy and Industrial Technology Development Organization (NEDO) and JSPS KAKENHI No. 18K18604.
\end{acknowledgments}

\appendix

\section{Brillouin--Wigner expansion}\label{BrillouinWignerExpansion}
In the main text, we obtained a formal solution of the generalized eigenvalue equation (\ref{EigenValueEquation}) using an expansion in the form the Lippmann--Schwinger equation. 
Here, we show that another formal solution can be obtained by using the Brillouin--Wigner expansion \cite{Ziman1969}. 

Equation (\ref{EigenValueEquation}) with respect to the $k$th eigenvalue $\lambda_{k} = 2s^{2}_{k}$ is the following eigenvalue equation:  
\begin{align}
2\mat{\mathsf{B}}\mat{\mathsf{B}}^{\top} \ket{\phi}^{\prime}_{k}
&= \lambda_{k} \ket{\phi}^{\prime}_{k}. \label{BrillouinWigner}
\end{align}
We note that the matrix on the left-hand side is 
\begin{align}
&2\mat{\mathsf{B}}\mat{\mathsf{B}}^{\top} 
= \mathcal{H}^{\prime}
= \mat{D}^{-1/2}_{U} \mathcal{H} \mat{D}^{-1/2}_{U}, \\
& \mathcal{H} \equiv -\mat{L} + 2 \mat{D}^{0}_{U} + 2\sum_{r} \frac{\ket{h}^{r}\ket{h}^{r \top}}{d^{v}_{r}}. 
\end{align}
In the absence of external hyperedges, we let 
\begin{align}
& \mathcal{H}^{\prime}_{0}
= \mat{D}^{0 \, -1/2}_{U} \mathcal{H}_{0} \mat{D}^{0 \, -1/2}_{U}, 
\hspace{20pt} 
\mathcal{H}_{0} \equiv -\mat{L} + 2 \mat{D}^{0}_{U}, 
\end{align}
and we define $\widetilde{\mathcal{H}}^{\prime} \equiv \mathcal{H}^{\prime} - \mathcal{H}^{\prime}_{0}$.

The eigenvalue equation (\ref{BrillouinWigner}) is then reformulated as 
\begin{align}
\left( \lambda_{k} - \mathcal{H}^{\prime}_{0} \right) \ket{\phi}^{\prime}_{k} = \widetilde{\mathcal{H}}^{\prime} \ket{\phi}^{\prime}_{k} \label{BrillouinWigner2}
\end{align}

In the Brillouin--Wigner expansion, we consider the following projection operators: 
\begin{align}
\Pi \equiv \ket{\varphi}^{\prime}_{k} \ket{\varphi}^{\prime \top}_{k}, \hspace{20pt} \Theta \equiv \mat{I} - \Pi, 
\end{align}
where $\ket{\varphi}^{\prime}_{k}$ ($= \mat{D}^{0 \, 1/2}_{U} \ket{\varphi}_{k}$) is the $k$th eigenvector of $\mathcal{H}^{\prime}_{0}$. 
In addition, ignoring the normalization of $\ket{\phi}^{\prime}_{k}$, we introduce a residual $\ket{\zeta}_{k}$ vector such that 
\begin{align}
\ket{\phi}^{\prime}_{k} = \ket{\varphi}^{\prime}_{k} + \ket{\zeta}_{k}, 
\hspace{20pt} \bra{\zeta}_{k} \ket{\varphi}^{\prime}_{k} = 0. \label{BrillouinWigner3}
\end{align}

By applying $\Theta$ from the left on the both sides of Eq.~(\ref{BrillouinWigner2}), we have 
\begin{align}
\Theta \left( \lambda_{k} - \mathcal{H}^{\prime}_{0} \right) \left( \ket{\varphi}^{\prime}_{k} + \ket{\zeta}_{k} \right)
&= \Theta \widetilde{\mathcal{H}}^{\prime} \ket{\phi}^{\prime}_{k} \notag\\
\left( \lambda_{k} - \lambda^{0}_{k} \right) \Theta \ket{\varphi}^{\prime}_{k} 
+ \left( \lambda_{k} - \mathcal{H}^{\prime}_{0} \right) \Theta \ket{\zeta}_{k}
&= \Theta \widetilde{\mathcal{H}}^{\prime} \ket{\phi}^{\prime}_{k} \notag\\
\left( \lambda_{k} - \mathcal{H}^{\prime}_{0} \right) \ket{\zeta}_{k} 
&= \Theta \widetilde{\mathcal{H}}^{\prime} \ket{\phi}^{\prime}_{k} \notag\\
\ket{\zeta}_{k} 
&= \left( \lambda_{k} - \mathcal{H}^{\prime}_{0} \right)^{-1} \Theta \widetilde{\mathcal{H}}^{\prime} \ket{\phi}^{\prime}_{k}. \label{BrillouinWigner4}
\end{align}
In the second line, we used $\mathcal{H}^{\prime}_{0} \ket{\varphi}^{\prime}_{k} = \lambda^{0}_{k} \ket{\varphi}^{\prime}_{k}$ and $\Theta \mathcal{H}^{\prime}_{0} \ket{\zeta}_{k} = \mathcal{H}^{\prime}_{0} \Theta \ket{\zeta}_{k}$ (because $\bra{\zeta}_{k} \ket{\varphi}^{\prime}_{k} = 0$). 
In the third line, we used $\Theta \ket{\varphi}^{\prime}_{k}=\ket{0}$ and $\Theta \ket{\zeta}_{k} = \ket{\zeta}_{k}$. 
In the fourth line, we defined $\left( \lambda_{k} - \mathcal{H}^{\prime}_{0} \right)^{-1}$ as the inverse of $\left( \lambda_{k} - \mathcal{H}^{\prime}_{0} \right)$. 
By substituting Eq.~(\ref{BrillouinWigner4}) into the first equation in Eq.~(\ref{BrillouinWigner3}), we have 
\begin{align}
\ket{\phi}^{\prime}_{k} 
&= \ket{\varphi}^{\prime}_{k} + \left( \lambda_{k} - \mathcal{H}^{\prime}_{0} \right)^{-1} \Theta \widetilde{\mathcal{H}}^{\prime} \ket{\phi}^{\prime}_{k} \notag\\
&= \ket{\varphi}^{\prime}_{k} 
+ \left( \lambda_{k} - \mathcal{H}^{\prime}_{0} \right)^{-1} \Theta \widetilde{\mathcal{H}}^{\prime} \ket{\varphi}^{\prime}_{k} \notag\\
&\hspace{10pt}+ \left( \lambda_{k} - \mathcal{H}^{\prime}_{0} \right)^{-1} \Theta \widetilde{\mathcal{H}}^{\prime} 
\left( \lambda_{k} - \mathcal{H}^{\prime}_{0} \right)^{-1} \Theta \widetilde{\mathcal{H}}^{\prime} \ket{\varphi}^{\prime}_{k} + \cdots. \label{BrillouinWigner5}
\end{align}
This formal solution is known as the Brillouin--Wigner expansion. 
Although this is similar to Eq.~(\ref{GF-3}), Eq.~(\ref{BrillouinWigner5}) is written in terms of the regularized vectors $\ket{\varphi}^{\prime}_{k}$ and $\ket{\phi}^{\prime}_{k}$. 
We note that both Eqs.~(\ref{GF-3}) and (\ref{BrillouinWigner5}) require $\lambda_{k}$ as well as $\lambda^{0}_{k}$.

\section{Derivation of the eigenvalue equation in the crude approximation}\label{sec:CrudeApproxDerivations}
\subsection{Original graph}
In the absence of external hyperedges, the generalized eigenvalue equation (\ref{EigenValueEquation}) is approximated as  
\begin{align}
\sum_{j \in U} A_{ij} \varphi_{kj} &= (\lambda^{0}_{k}-1) d^{u}_{i} \varphi_{ki}, \notag\\
\frac{1}{N_{\sigma}} \sum_{i \in U_{\sigma}} \sum_{j \in U} A_{ij} \varphi_{kj} 
&= \frac{\lambda^{0}_{k}-1}{N_{\sigma}} \sum_{i \in U_{\sigma}} d^{u}_{i} \varphi_{ki} 
\approx (\lambda^{0}_{k}-1) c_{\sigma} \bar{\varphi}_{k \sigma}. \label{CrudeApproximationRight}
\end{align}
The left-hand side is approximated as  
\begin{align}
\frac{1}{N_{\sigma}} \sum_{i \in U_{\sigma}} \sum_{j \in U} A_{ij} \varphi_{kj} 
&\approx \frac{1}{N_{\sigma}} \sum_{\sigma^{\prime}} \left( \sum_{i \in U_{\sigma}} \sum_{j \in U_{\sigma^{\prime}}} A_{ij} \right) \bar{\varphi}_{k \sigma^{\prime}} \notag\\
&= \frac{1}{N_{\sigma}} \sum_{\sigma^{\prime}} \left(1 + \delta(\sigma, \sigma^{\prime}) \right) e_{\sigma \sigma^{\prime}} \bar{\varphi}_{k \sigma^{\prime}} \notag\\
&= c_{\sigma} \sum_{\sigma^{\prime}} f_{\sigma \sigma^{\prime}} \bar{\varphi}_{k \sigma^{\prime}}. \label{CrudeApproximationLeft}
\end{align}
Then, we obtain the approximated generalized eigenvalue equation (\ref{CrudeApproxUnperturbed}) in the main text. 

\subsection{Scotch-taped graph}
In the presence of external hyperedges, the element-wise expression of the generalized eigenvalue equation (\ref{EigenValueEquation}) is 
\begin{align}
& \sum_{j \in U} \left( A_{ij} + 2\sum_{r} \frac{h^{r}_{i} h^{r}_{j} }{d^{v}_{r}} \right) \phi_{kj} 
= \left( d^{u}_{i} (\lambda_{k}-1) + \lambda_{k} \sum_{r} h^{r}_{i} \right) \phi_{ki}. \label{GeneralizedEigenvalueEquation}
\end{align}
By the definition of $\overline{h^{r}_{\sigma}}$, 
\begin{align}
& \frac{1}{N_{\sigma}} \sum_{i \in U_{\sigma}} \sum_{j \in U} 2\sum_{r} \frac{h^{r}_{i} h^{r}_{j} }{d^{v}_{r}} \phi_{kj} \notag\\
&\hspace{10pt}= 2\sum_{r} \frac{\overline{h^{r}_{\sigma}} }{ \sum_{\sigma^{\prime\prime}} \overline{h^{r}_{\sigma^{\prime\prime}}} N_{\sigma^{\prime\prime}} } \sum_{j \in U} h^{r}_{j} \phi_{kj} \notag\\
&\hspace{10pt}\approx 2\sum_{r} \frac{\overline{h^{r}_{\sigma}} }{ \sum_{\sigma^{\prime\prime}} \overline{h^{r}_{\sigma^{\prime\prime}}} N_{\sigma^{\prime\prime}} } \sum_{\sigma^{\prime}} \overline{h^{r}_{\sigma^{\prime}}} N_{\sigma^{\prime}} \phi_{k \sigma^{\prime}}. \label{CrudeApproximationLeft2}
\end{align}
Here, we used $d^{v}_{r} = \sum_{\sigma^{\prime\prime}} \overline{h^{r}_{\sigma^{\prime\prime}}} N_{\sigma^{\prime\prime}}$. 
Using Eqs.~(\ref{CrudeApproximationRight}), (\ref{CrudeApproximationLeft}), and (\ref{CrudeApproximationLeft2}), Eq.~(\ref{GeneralizedEigenvalueEquation}) is approximated as Eq.~(\ref{CrudeApproxPerturbed}) in the main text. 

\begin{widetext}

\section{Replica method}\label{sec:ReplicaMethod}
In this section, we derive the message-passing equation of the second-largest eigenvector-element distribution for the microcanonical SBM. 
Equation (\ref{PhysicalCavityEq}) corresponds to Eq.~(\ref{PhysicalCavityEqMain}) in the main text; $A_{Q}$ and $H_{Q}$ in Eq.~(\ref{PhysicalCavityEq}) are replaced by $\mathsf{A}$ and $\mathsf{H}$ in Eq.~(\ref{PhysicalCavityEqMain}) in the main text, respectively.

To calculate $\left[ \lambda_{2} \right]_{\mat{B}^{0}}$ in Eq.~(\ref{ReplicaEqn}), we should calculate the moment $\left[ Z^{n}(\beta, \lambda, \gamma) \right]_{\mat{B}^{0}}$. 
According to Eqs.~(\ref{PartitionFunction}) and (\ref{EnergyFunction}), the $n$th power of $Z^{n}(\beta, \mu, \lambda, \gamma)$ is 
\begin{align}
Z^{n}(\beta, \mu, \lambda, \gamma)
&= 
\int \prod_{i \in U} \prod_{a=1}^{n} dx_{ia} \, 
\exp\Biggl( \beta \sum_{a=1}^{n} \biggl( 
\frac{1}{2} \lambda N 
+ \frac{1}{2} \left( \sum_{i \in U} x_{ia} \mat{B}^{0}_{i\alpha} \right)^{2} + \sum_{r=1}^{R} \frac{1}{d^{v}_{r}} \left(\sum_{i \in U} x_{ia} h^{r}_{i}\right)^{2} \notag\\
&\hspace{100pt}- \frac{\lambda}{2} \sum_{i \in U} d^{u}_{i} x^{2}_{ia} 
- \gamma \sum_{i \in U} d^{u}_{i} x_{ia} 
\biggr) \Biggr).
\end{align}
By introducing the auxiliary variables 
\begin{align}
& y_{\alpha a} = \sum_{i \in U} x_{ia} \mat{B}^{0}_{i\alpha},  \hspace{10pt} (\alpha \in V^{0}) \\
& m_{ra} = \frac{1}{N} \sum_{i \in U} x_{ia} h^{r}_{i}, 
\end{align}
we have 
\begin{align}
Z^{n}(\beta, \lambda, \gamma)
&= 
\int \prod_{i,a} dx_{ia} \, 
\int \prod_{\alpha,a} dy_{\alpha a} 
\int \prod_{r,a} dm_{r a} \prod_{\alpha \in V^{0}} \delta\left( y_{\alpha a} - \sum_{i \in U} x_{ia} \mat{B}^{0}_{i\alpha} \right) 
\prod_{\alpha \in V^{0}} \delta\left( m_{ra} - \frac{1}{N}\sum_{i \in U} x_{ia} h^{r}_{i} \right)\notag\\
& \times \exp \Biggl( 
\frac{1}{2} n \beta \lambda N
+\frac{\beta}{2} \sum_{a=1}^{n} \sum_{\alpha \in V^{0}} y_{\alpha a}^{2} 
+ \beta N^{2} \sum_{a=1}^{n} \sum_{r=1}^{R} \frac{1}{d^{v}_{r}} m_{ra}^{2} 
- \beta \sum_{a=1}^{n} \sum_{i \in U} d^{u}_{i} (\frac{\lambda}{2}  x^{2}_{ia} + \gamma x_{ia})
\Biggr) \\
&= 
\int \prod_{i,a} dx_{ia} \, 
\int \prod_{\alpha,a} \frac{d\hat{y}_{\alpha a} dy_{\alpha a}}{2\pi} 
\int \prod_{r,a} \frac{\beta N d\hat{m}_{r a} dm_{r a}}{2\pi} \notag\\
& \times \exp \Biggl( 
\frac{1}{2} n \beta \lambda N
+\frac{\beta}{2} \sum_{a=1}^{n} \sum_{\alpha \in V^{0}} y_{\alpha a}^{2} 
+ \beta N^{2} \sum_{a=1}^{n} \sum_{r=1}^{R} \frac{1}{d^{v}_{r}} m_{ra}^{2} \notag\\
&- \beta \sum_{a=1}^{n} \sum_{i \in U} \sum_{\alpha \in V_{0}} B^{0}_{i\alpha} (\frac{\lambda}{2}  x^{2}_{ia} + \gamma x_{ia})
- \beta \sum_{a=1}^{n} \sum_{i \in U}  \sum_{r} h^{r}_{i} (\frac{\lambda}{2}  x^{2}_{ia} + \gamma x_{ia})\notag\\
&- \beta \sum_{a} \sum_{\alpha \in V^{0}} \hat{y}_{\alpha a} \left( y_{\alpha a} - \sum_{i \in U} x_{ia} \mat{B}^{0}_{i\alpha} \right) 
- \beta \sum_{a} \sum_{r} \hat{m}_{ra} \left( m_{ra}N - \sum_{i \in U} x_{ia} h^{r}_{i} \right)
\Biggr).
\end{align}
Here, we used the fact that the degree $d^{u}_{i}$ of a physical node is decomposed as $ d^{u}_{i} = \sum_{\alpha \in V_{0}} B^{0}_{i\alpha} + \sum_{r} h^{r}_{i}$. 
Then, the moment can be calculated as follows: 
\begin{align}
& \left[ Z^{n}(\beta, \lambda, \gamma) \right]_{\mat{B}^{0}} \notag\\
&= 
\int \prod_{i,a} dx_{ia} \, 
\int \prod_{\alpha,a} \frac{d\hat{y}_{\alpha a} dy_{\alpha a}}{2\pi} 
\int \prod_{r,a} \frac{\beta N d\hat{m}_{r a} dm_{r a}}{2\pi} \, 
\exp \Biggl( \frac{1}{2} n \beta \lambda N 
- \beta \sum_{a=1}^{n} \sum_{i \in U}  \sum_{r} h^{r}_{i} (\frac{\lambda}{2} x^{2}_{ia} + \gamma x_{ia}) \notag\\
&\hspace{20pt} +\frac{\beta}{2} \sum_{a=1}^{n} \sum_{\alpha \in V^{0}} \left( y_{\alpha a}^{2} - 2 \hat{y}_{\alpha a} y_{\alpha a}\right)
+ \beta N \sum_{a=1}^{n} \sum_{r=1}^{R} \left( \frac{N}{d^{v}_{r}} m_{ra}^{2} - \hat{m}_{ra} m_{ra} \right) 
+ \beta \sum_{a} \sum_{i \in U} \sum_{r} \hat{m}_{ra} x_{ia} h^{r}_{i} \Biggr) \notag\\
&\hspace{20pt} \times \left[ \exp\left( -\beta \sum_{a} \sum_{\alpha \in V^{0}} \sum_{i \in U} 
\mat{B}^{0}_{i\alpha} (\frac{\lambda}{2} x^{2}_{ia} + \gamma x_{ia} - x_{ia} \hat{y}_{\alpha a}) \right) \right]_{\mat{B}^{0}}.
\end{align}
By using the probability distribution of the microcanonical SBM (Eq.~(\ref{microcanonicalSBM})), the last factor is calculated as 
\begin{align}
& \left[ \exp\left( -\beta \sum_{a} \sum_{\alpha \in V^{0}} \sum_{i \in U} 
\mat{B}^{0}_{i\alpha} (\frac{\lambda}{2} x^{2}_{ia} + \gamma x_{ia} - x_{ia} \hat{y}_{\alpha a}) \right) \right]_{\mat{B}^{0}} \notag\\
&= \frac{1}{\mathcal{N}_{G}} \sum_{\{ B^{0}_{i\alpha} \}} 
\prod_{\sigma} \prod_{\alpha \in V_{\sigma \sigma}} \delta\left( \sum_{i \in U_{\sigma}}B^{0}_{i \alpha}, 2 \right)
\prod_{\sigma < \sigma^{\prime}} \prod_{\alpha \in V^{0}_{\sigma \sigma^{\prime}}} \delta\left( \sum_{i \in U_{\sigma}}B^{0}_{i \alpha}, 1 \right) \delta\left( \sum_{j \in U_{\sigma^{\prime}}}B^{0}_{j \alpha}, 1 \right) \notag\\
& \hspace{80pt}\times\prod_{\alpha \in V^{0}} \exp\left( -\beta \sum_{a} \sum_{i \in U} 
\mat{B}^{0}_{i\alpha} (\frac{\lambda}{2}  x^{2}_{ia} + \gamma x_{ia} - x_{ia} \hat{y}_{\alpha a}) \right) \\
&= \frac{1}{\mathcal{N}_{G}} \sum_{\{ B^{0}_{i\alpha} \}} 
\oint \prod_{\sigma=1}^{K} \prod_{\alpha \in V^{0}_{\sigma \sigma}} \frac{dz_{\alpha \sigma}}{2\pi \imgunit} z_{\alpha \sigma}^{(1-\sum_{i \in U_{\sigma}}B^{0}_{i\alpha})} \notag\\
&\hspace{40pt}\times \oint \prod_{\sigma < \sigma^{\prime}} \prod_{\alpha \in V^{0}_{\sigma \sigma^{\prime}}} 
\frac{dz_{\alpha \sigma}}{2\pi \imgunit} \frac{dz_{\alpha \sigma^{\prime}}}{2\pi \imgunit} 
z_{\alpha}^{-\sum_{i \in U_{\sigma}}B^{0}_{i\alpha}} z_{\alpha}^{-\sum_{j \in U_{\sigma^{\prime}}}B^{0}_{j\alpha}} \notag\\
& \hspace{40pt}\times\prod_{\alpha \in V^{0}} \exp\left( -\beta \sum_{a} \sum_{i \in U} 
\mat{B}^{0}_{i\alpha} (\frac{\lambda}{2}  x^{2}_{ia} + \gamma x_{ia} - x_{ia} \hat{y}_{\alpha a}) \right) \\
&= \frac{1}{\mathcal{N}_{G}} 
\prod_{\sigma=1}^{K} \left( 
\sum_{\{ B^{0}_{i\alpha} \}^{i \in U_{\sigma}}_{\alpha \in V^{0}_{\sigma \sigma}}} 
\oint \prod_{\alpha \in V^{0}_{\sigma \sigma}} \left( \frac{dz_{\alpha \sigma}}{2\pi \imgunit} z_{\alpha \sigma} 
\prod_{i \in U_{\sigma}} z_{\alpha \sigma}^{-B^{0}_{i \alpha}} 
\prod_{i \in U_{\sigma}} 
\mathrm{e}^{-\beta \sum_{a} \mat{B}^{0}_{i\alpha} (\frac{\lambda}{2}  x^{2}_{ia} + \gamma x_{ia} - x_{ia} \hat{y}_{\alpha a})}
\right)
\right) \notag\\
& \times \prod_{\sigma < \sigma^{\prime}}^{K} \left( 
\sum_{\{ B^{0}_{i\alpha} \}^{i \in U_{\sigma} \cup U_{\sigma^{\prime}}}_{\alpha \in V^{0}_{\sigma \sigma^{\prime}}}} 
\oint \prod_{\alpha \in V^{0}_{\sigma \sigma^{\prime}}} \left( \frac{dz_{\alpha \sigma} dz_{\alpha \sigma^{\prime}}}{(2\pi \imgunit)^{2}} 
\prod_{i \in U_{\sigma}} z_{\alpha \sigma}^{-B^{0}_{i \alpha}} \prod_{j \in U_{\sigma^{\prime}}} z_{\alpha \sigma^{\prime}}^{-B^{0}_{j \alpha}} 
\prod_{i \in U_{\sigma}\cup U_{\sigma^{\prime}}} 
\mathrm{e}^{-\beta \sum_{a} \mat{B}^{0}_{i\alpha} (\frac{\lambda}{2}  x^{2}_{ia} + \gamma x_{ia} - x_{ia} \hat{y}_{\alpha a})}
\right) 
\right). 
\end{align}

Here, we introduce the following order-parameter functions: 
\begin{align}
& Q_{\sigma}(\ket{\mathsf{x}}) \equiv \frac{1}{N_{\sigma}} \sum_{i \in U_{\sigma}} \prod_{a=1}^{n} \delta\left( \mathsf{x}_{a} - x_{ia} \right), \\
& P_{\sigma \sigma^{\prime}}(\ket{\mathsf{y}}) \equiv \frac{1}{e_{\sigma \sigma^{\prime}}} \sum_{\alpha \in V^{0}_{\sigma \sigma^{\prime}}} z_{\alpha \sigma}^{-1} \prod_{a=1}^{n} \delta\left( \mathsf{y}_{a} - \hat{y}_{\alpha a} \right).
\end{align}
$P_{\sigma \sigma^{\prime}}(\ket{\mathsf{y}})$ is asymmetric with respect to $\sigma$ and $\sigma^{\prime}$, and it is order-sensitive. 
By using $Q_{\sigma}(\ket{\mathsf{x}})$ and $P_{\sigma \sigma^{\prime}}(\ket{\mathsf{y}})$, 
for the factor of $V^{0}_{\sigma \sigma}$, 
\begin{align}
& \sum_{\{ B^{0}_{i\alpha} \}^{i \in U_{\sigma}}_{\alpha \in V^{0}_{\sigma \sigma}}} 
\oint \prod_{\alpha \in V^{0}_{\sigma \sigma}} \left( \frac{dz_{\alpha \sigma}}{2\pi \imgunit} z_{\alpha \sigma} 
\prod_{i \in U_{\sigma}} z_{\alpha \sigma}^{-B^{0}_{i \alpha}} 
\prod_{i \in U_{\sigma}} 
\mathrm{e}^{-\beta \sum_{a} \mat{B}^{0}_{i\alpha} (\frac{\lambda}{2}  x^{2}_{ia} + \gamma x_{ia} - x_{ia} \hat{y}_{\alpha a})} \right) \notag\\
&= 
\oint \prod_{\alpha \in V^{0}_{\sigma \sigma}} \left( \frac{dz_{\alpha \sigma}}{2\pi \imgunit} z_{\alpha \sigma} \right)
\prod_{i \in U_{\sigma}} \prod_{\alpha \in V^{0}_{\sigma \sigma}} \left( 1 + z_{\alpha \sigma}^{-1} 
\mathrm{e}^{-\beta \sum_{a} (\frac{\lambda}{2}  x^{2}_{ia} + \gamma x_{ia} - x_{ia} \hat{y}_{\alpha a})} \right) \notag\\
&\approx 
\oint \prod_{\alpha \in V^{0}_{\sigma \sigma}} \left( \frac{dz_{\alpha \sigma}}{2\pi \imgunit} z_{\alpha \sigma} \right)
\exp \left( \sum_{i \in U_{\sigma}} \sum_{\alpha \in V^{0}_{\sigma \sigma}} z_{\alpha \sigma}^{-1} \mathrm{e}^{-\beta \sum_{a} (\frac{\lambda}{2}  x^{2}_{ia} + \gamma x_{ia} - x_{ia} \hat{y}_{\alpha a})} \right) \notag\\
&= \oint \prod_{\alpha \in V^{0}_{\sigma \sigma}} \left( \frac{dz_{\alpha \sigma}}{2\pi \imgunit} z_{\alpha \sigma} \right) 
\exp\left( N_{\sigma} e_{\sigma \sigma} \int d\ket{\mathsf{x}} d\ket{\mathsf{y}} \, Q_{\sigma}(\ket{\mathsf{x}}) P_{\sigma \sigma}(\ket{\mathsf{y}}) 
\mathrm{e}^{ -\beta \sum_{a} (\frac{\lambda}{2}  \mathsf{x}^{2}_{a} + \gamma \mathsf{x}_{a} - \mathsf{x}_{a} \mathsf{y}_{a}) } \right),
\end{align}
where we assumed that $z_{\alpha \sigma}$ is sufficiently large. 
Similarly, for the factor of $V^{0}_{\sigma \sigma^{\prime}}$ ($\sigma < \sigma^{\prime}$), 
\begin{align}
& \sum_{\{ B^{0}_{i\alpha} \}^{i \in U_{\sigma} \cup U_{\sigma^{\prime}}}_{\alpha \in V^{0}_{\sigma \sigma^{\prime}}}} 
\oint \prod_{\alpha \in V^{0}_{\sigma \sigma^{\prime}}} \left( \frac{dz_{\alpha \sigma} dz_{\alpha \sigma^{\prime}}}{(2\pi \imgunit)^{2}} 
\prod_{i \in U_{\sigma}} z_{\alpha \sigma}^{-B^{0}_{i \alpha}} \prod_{j \in U_{\sigma^{\prime}}} z_{\alpha \sigma^{\prime}}^{-B^{0}_{j \alpha}} 
\prod_{i \in U_{\sigma}\cup U_{\sigma^{\prime}}} 
\mathrm{e}^{-\beta \sum_{a} \mat{B}^{0}_{i\alpha} (\frac{\lambda}{2}  x^{2}_{ia} + \gamma x_{ia} - x_{ia} \hat{y}_{\alpha a})}
\right) \notag\\
&= 
\prod_{\alpha \in V^{0}_{\sigma \sigma^{\prime}}} \left( \oint \frac{dz_{\alpha \sigma}}{2\pi \imgunit} \prod_{i \in U_{\sigma}} 
\left( \sum_{B^{0}_{i \alpha}} z_{\alpha \sigma}^{-B^{0}_{i \alpha}} \mathrm{e}^{ -\beta \sum_{a} \mat{B}^{0}_{i\alpha} (\frac{\lambda}{2}  x^{2}_{ia} + \gamma x_{ia} - x_{ia} \hat{y}_{\alpha a}) } \right) \right) \notag\\
&\hspace{20pt}\times 
\prod_{\alpha \in V^{0}_{\sigma \sigma^{\prime}}} \left( \oint \frac{dz_{\alpha \sigma^{\prime}}}{2\pi \imgunit} \prod_{j \in U_{\sigma^{\prime}}} 
\left( \sum_{B^{0}_{j \alpha}} z_{\alpha \sigma^{\prime}}^{-B^{0}_{j \alpha}} \mathrm{e}^{ -\beta \sum_{a} \mat{B}^{0}_{j\alpha} (\frac{\lambda}{2}  x^{2}_{ja} + \gamma x_{ja} - x_{ja} \hat{y}_{\alpha a}) } \right) \right) \notag\\
&\approx 
\oint \prod_{\alpha \in V^{0}_{\sigma \sigma^{\prime}}} \frac{dz_{\alpha \sigma}}{2\pi \imgunit} \frac{dz_{\alpha \sigma^{\prime}}}{2\pi \imgunit} 
\exp\biggl( 
N_{\sigma} e_{\sigma \sigma^{\prime}} \int d\ket{\mathsf{x}} d\ket{\mathsf{y}} \, Q_{\sigma}(\ket{\mathsf{x}}) P_{\sigma \sigma^{\prime}}(\ket{\mathsf{y}}) \mathrm{e}^{-\beta \sum_{a} (\frac{\lambda}{2}  \mathsf{x}^{2}_{a} + \gamma \mathsf{x}_{a} - \mathsf{x}_{a} \mathsf{y}_{a})} \notag\\
&\hspace{80pt}+ N_{\sigma^{\prime}} e_{\sigma \sigma^{\prime}} \int d\ket{\mathsf{x}} d\ket{\mathsf{y}} \, Q_{\sigma^{\prime}}(\ket{\mathsf{x}}) P_{\sigma^{\prime} \sigma}(\ket{\mathsf{y}}) \mathrm{e}^{-\beta \sum_{a} (\frac{\lambda}{2}  \mathsf{x}^{2}_{a} + \gamma \mathsf{x}_{a} - \mathsf{x}_{a} \mathsf{y}_{a})}. 
\biggr)
\end{align}
Then, 
\begin{align}
& \left[ \exp\left( -\beta \sum_{a} \sum_{\alpha \in V^{0}} \sum_{i \in U} 
\mat{B}^{0}_{i\alpha} (\frac{\lambda}{2}  x^{2}_{ia} + \gamma x_{ia} - x_{ia} \hat{y}_{\alpha a}) \right) \right]_{\mat{B}^{0}} \notag\\
&= \frac{1}{\mathcal{N}_{G}} 
\oint \prod_{\sigma} \prod_{\alpha \in V^{0}_{\sigma \sigma}} \frac{dz_{\alpha \sigma}}{2\pi \imgunit} z_{\alpha \sigma}  
\oint \prod_{\sigma < \sigma^{\prime}} \prod_{\alpha \in V^{0}_{\sigma \sigma^{\prime}}} \frac{dz_{\alpha \sigma}}{2\pi \imgunit} \frac{dz_{\alpha \sigma^{\prime}}}{2\pi \imgunit} \notag\\
&\hspace{10pt} \times \int \prod_{\sigma} N_{\sigma} dQ_{\sigma}(\ket{\mathsf{x}}) \delta\left( N_{\sigma} Q_{\sigma}(\ket{\mathsf{x}}) - \sum_{i \in U_{\sigma}} \prod_{a=1}^{n} \delta\left( \mathsf{x}_{a} - x_{ia} \right) \right) \notag\\
&\hspace{10pt} \times \int \prod_{\sigma, \sigma^{\prime}} e_{\sigma \sigma^{\prime}} dP_{\sigma \sigma^{\prime}}(\ket{\mathsf{y}}) \delta\left( e_{\sigma \sigma^{\prime}} P_{\sigma \sigma^{\prime}}(\ket{\mathsf{y}}) - \sum_{\alpha \in V^{0}_{\sigma \sigma^{\prime}}} z_{\alpha \sigma}^{-1} \prod_{a=1}^{n} \delta\left( \mathsf{y}_{a} - \hat{y}_{\alpha a} \right) \right) \notag\\
&\hspace{10pt}\times \exp\left( \sum_{\sigma, \sigma^{\prime}} N_{\sigma} e_{\sigma \sigma^{\prime}} \int d\ket{\mathsf{x}} d\ket{\mathsf{y}} \, Q_{\sigma}(\ket{\mathsf{x}}) P_{\sigma \sigma^{\prime}}(\ket{\mathsf{y}}) 
\mathrm{e}^{ -\beta \sum_{a} (\frac{\lambda}{2}  \mathsf{x}^{2}_{a} + \gamma \mathsf{x}_{a} - \mathsf{x}_{a} \mathsf{y}_{a}) } \right) \notag\\
&= \frac{1}{\mathcal{N}_{G}} 
\oint \prod_{\sigma} \prod_{\alpha \in V^{0}_{\sigma \sigma}} \frac{dz_{\alpha \sigma}}{2\pi \imgunit} z_{\alpha \sigma}  
\oint \prod_{\sigma < \sigma^{\prime}} \prod_{\alpha \in V^{0}_{\sigma \sigma^{\prime}}} \frac{dz_{\alpha \sigma}}{2\pi \imgunit} \frac{dz_{\alpha \sigma^{\prime}}}{2\pi \imgunit} 
\int \prod_{\sigma} N_{\sigma} \frac{d\hat{Q}_{\sigma}(\ket{\mathsf{x}}) dQ_{\sigma}(\ket{\mathsf{x}})}{2\pi} 
\int \prod_{\sigma, \sigma^{\prime}} e_{\sigma \sigma^{\prime}} \frac{d\hat{P}_{\sigma \sigma^{\prime}}(\ket{\mathsf{y}}) dP_{\sigma \sigma^{\prime}}(\ket{\mathsf{y}})}{2\pi} \notag\\
&\hspace{10pt} \times \exp \Biggl( 
-\sum_{\sigma} N_{\sigma} \int d\ket{\mathsf{x}} \, \hat{Q}_{\sigma}(\ket{\mathsf{x}}) Q_{\sigma}(\ket{\mathsf{x}})
-\sum_{\sigma \sigma^{\prime}} e_{\sigma \sigma^{\prime}} \int d\ket{\mathsf{y}} \, \hat{P}_{\sigma \sigma^{\prime}}(\ket{\mathsf{y}}) P_{\sigma \sigma^{\prime}}(\ket{\mathsf{y}}) \notag\\
&\hspace{40pt} +\sum_{\sigma, \sigma^{\prime}} N_{\sigma} e_{\sigma \sigma^{\prime}} \int d\ket{\mathsf{x}} d\ket{\mathsf{y}} \, Q_{\sigma}(\ket{\mathsf{x}}) P_{\sigma \sigma^{\prime}}(\ket{\mathsf{y}}) \mathrm{e}^{-\beta \sum_{a} (\frac{\lambda}{2}  \mathsf{x}^{2}_{a} + \gamma \mathsf{x}_{a} - \mathsf{x}_{a} \mathsf{y}_{a})} 
+ \sum_{\sigma} \sum_{i \in U_{\sigma}} \hat{Q}_{\sigma}(\ket{x}_{i}) 
+ \sum_{\sigma \sigma^{\prime}} \sum_{\alpha \in V^{0}_{\sigma \sigma^{\prime}}} z_{\alpha \sigma}^{-1} \hat{P}_{\sigma \sigma^{\prime}}(\ket{\hat{y}}_{\alpha})
\Biggr). 
\end{align}
The integrals with respect to $z_{\alpha \sigma}$ for $\alpha \in V^{0}_{\sigma \sigma}$ and $\alpha \in V^{0}_{\sigma \sigma^{\prime}} \, (\sigma < \sigma^{\prime})$ are calculated as 
\begin{align}
& \oint \frac{dz_{\alpha \sigma}}{2\pi \imgunit} z_{\alpha \sigma} 
\exp\left( z_{\alpha \sigma}^{-1} \hat{P}_{\sigma \sigma}(\ket{\hat{y}}_{\alpha}) \right) 
= \oint \frac{dz_{\alpha \sigma}}{2\pi \imgunit} z_{\alpha \sigma} 
\sum_{k=0}^{\infty} \frac{1}{k!} \left( z_{\alpha \sigma}^{-1} \hat{P}_{\sigma \sigma}(\ket{\hat{y}}_{\alpha}) \right)^{k} 
= \frac{1}{2}\hat{P}^{2}_{\sigma \sigma}(\ket{\hat{y}}_{\alpha}), \\
& \oint \frac{dz_{\alpha \sigma}}{2\pi \imgunit} \frac{dz_{\alpha \sigma^{\prime}}}{2\pi \imgunit} 
\exp\left( z_{\alpha \sigma}^{-1} \hat{P}_{\sigma \sigma^{\prime}}(\ket{\hat{y}}_{\alpha}) 
+ z_{\alpha \sigma^{\prime}}^{-1} \hat{P}_{\sigma^{\prime} \sigma}(\ket{\hat{y}}_{\alpha}) \right) 
= \hat{P}_{\sigma \sigma^{\prime}}(\ket{\hat{y}}_{\alpha}) \hat{P}_{\sigma^{\prime} \sigma}(\ket{\hat{y}}_{\alpha}),
\end{align}
respectively. Therefore, we obtain 
\begin{align}
& \left[ \exp\left( -\beta \sum_{a} \sum_{\alpha \in V^{0}} \sum_{i \in U} 
\mat{B}^{0}_{i\alpha} (\frac{\lambda}{2}  x^{2}_{ia} + \gamma x_{ia} - x_{ia} \hat{y}_{\alpha a}) \right) \right]_{\mat{B}^{0}} \notag\\
&= \frac{1}{\mathcal{N}_{G}} 
\int \prod_{\sigma} N_{\sigma} \frac{d\hat{Q}_{\sigma}(\ket{\mathsf{x}}) dQ_{\sigma}(\ket{\mathsf{x}})}{2\pi} 
\int \prod_{\sigma, \sigma^{\prime}} e_{\sigma \sigma^{\prime}} \frac{d\hat{P}_{\sigma \sigma^{\prime}}(\ket{\mathsf{y}}) dP_{\sigma \sigma^{\prime}}(\ket{\mathsf{y}})}{2\pi} \notag\\
&\hspace{10pt} \times \exp \Biggl( 
-\sum_{\sigma} N_{\sigma} \int d\ket{\mathsf{x}} \, \hat{Q}_{\sigma}(\ket{\mathsf{x}}) Q_{\sigma}(\ket{\mathsf{x}})
-\sum_{\sigma \sigma^{\prime}} e_{\sigma \sigma^{\prime}} \int d\ket{\mathsf{y}} \, \hat{P}_{\sigma \sigma^{\prime}}(\ket{\mathsf{y}}) P_{\sigma \sigma^{\prime}}(\ket{\mathsf{y}}) \notag\\
&\hspace{40pt} +\sum_{\sigma, \sigma^{\prime}} N_{\sigma} e_{\sigma \sigma^{\prime}} \int d\ket{\mathsf{x}} d\ket{\mathsf{y}} \, Q_{\sigma}(\ket{\mathsf{x}}) P_{\sigma \sigma^{\prime}}(\ket{\mathsf{y}}) \mathrm{e}^{-\beta \sum_{a} (\frac{\lambda}{2}  \mathsf{x}^{2}_{a} + \gamma \mathsf{x}_{a} - \mathsf{x}_{a} \mathsf{y}_{a})} \notag\\
&\hspace{40pt} + \sum_{\sigma} \sum_{i \in U_{\sigma}} \hat{Q}_{\sigma}(\ket{x}_{i}) 
+ \sum_{\sigma} \sum_{\alpha \in V^{0}_{\sigma \sigma}} \log\left( \frac{1}{2}\hat{P}^{2}_{\sigma \sigma}(\ket{\hat{y}}_{\alpha}) \right)
+ \sum_{\sigma < \sigma^{\prime}} \sum_{\alpha \in V^{0}_{\sigma \sigma^{\prime}}} \log\left( \hat{P}_{\sigma \sigma^{\prime}}(\ket{\hat{y}}_{\alpha}) \hat{P}_{\sigma^{\prime} \sigma}(\ket{\hat{y}}_{\alpha}) \right)
\Biggr). 
\end{align}

The overall moment is 
\begin{align}
\left[ Z^{n}(\beta, \lambda, \gamma) \right]_{\mat{B}^{0}} 
&= \frac{1}{\mathcal{N}_{G}} 
\int \prod_{r,a} \frac{\beta N d\hat{m}_{r a} dm_{r a}}{2\pi} \, 
\int \prod_{\sigma} N_{\sigma} \frac{d\hat{Q}_{\sigma}(\ket{\mathsf{x}}) dQ_{\sigma}(\ket{\mathsf{x}})}{2\pi} 
\int \prod_{\sigma, \sigma^{\prime}} e_{\sigma \sigma^{\prime}} \frac{d\hat{P}_{\sigma \sigma^{\prime}}(\ket{\mathsf{y}}) dP_{\sigma \sigma^{\prime}}(\ket{\mathsf{y}})}{2\pi} \notag\\
&\hspace{10pt}\times \exp \Biggl( \frac{1}{2} n \beta \lambda N 
+ \beta N \sum_{a=1}^{n} \sum_{r=1}^{R} \left( \frac{N}{d^{v}_{r}} m_{ra}^{2} - \hat{m}_{ra} m_{ra} \right) \notag\\
&\hspace{10pt}+ \sum_{\sigma} \sum_{i \in U_{\sigma}} \log \Phi_{i}(n, \beta, \lambda, \gamma, \{\hat{m}_{r}\}) 
+ \sum_{\sigma \le \sigma^{\prime}} \sum_{\alpha \in V_{\sigma \sigma^{\prime}}} \log \Psi_{\alpha}(n, \beta) 
+ \Xi(n, \beta, \lambda, \gamma)
\Biggr), \label{ZnMoment}
\end{align}
where 
\begin{align}
& \Phi_{i}(n, \beta, \lambda, \gamma, \{\hat{m}_{r}\}) = 
\int \prod_{a} dx_{ia} \, 
\exp\left( 
- \beta \sum_{a=1}^{n} \sum_{r} h^{r}_{i} (\frac{\lambda}{2}  x^{2}_{ia} + \gamma x_{ia} - \hat{m}_{ra} x_{ia})
\right) 
\mathrm{e}^{\hat{Q}_{\sigma}(\ket{x}_{i})} \notag\\
&\hspace{300pt} \text{for } i \in U_{\sigma}, \\
& \Psi_{\alpha}(n, \beta) = 
\frac{1}{1+\delta(\sigma, \sigma^{\prime})} 
\int \prod_{a} \frac{d\hat{y}_{\alpha a} dy_{\alpha a}}{2\pi} \, \hat{P}_{\sigma \sigma^{\prime}} (\ket{\hat{y}}_{\alpha}) \hat{P}_{\sigma^{\prime} \sigma} (\ket{\hat{y}}_{\alpha}) 
\exp\left( \frac{\beta}{2} \sum_{a=1}^{n} \left( y_{\alpha a}^{2} - 2 \hat{y}_{\alpha a} y_{\alpha a}\right) \right) \notag\\
&\hspace{300pt} \text{for } \alpha \in V^{0}_{\sigma \sigma^{\prime}} \, (\sigma \le \sigma^{\prime}), \\
& \Xi(n, \beta, \lambda, \gamma) = -\sum_{\sigma} N_{\sigma} \int d\ket{\mathsf{x}} \, \hat{Q}_{\sigma}(\ket{\mathsf{x}}) Q_{\sigma}(\ket{\mathsf{x}})
-\sum_{\sigma \sigma^{\prime}} e_{\sigma \sigma^{\prime}} \int d\ket{\mathsf{y}} \, \hat{P}_{\sigma \sigma^{\prime}}(\ket{\mathsf{y}}) P_{\sigma \sigma^{\prime}}(\ket{\mathsf{y}}) \notag\\
&\hspace{60pt}+\sum_{\sigma, \sigma^{\prime}} N_{\sigma} e_{\sigma \sigma^{\prime}} \int d\ket{\mathsf{x}} d\ket{\mathsf{y}} \, Q_{\sigma}(\ket{\mathsf{x}}) P_{\sigma \sigma^{\prime}}(\ket{\mathsf{y}}) \mathrm{e}^{-\beta \sum_{a} (\frac{\lambda}{2}  \mathsf{x}^{2}_{a} + \gamma \mathsf{x}_{a} - \mathsf{x}_{a} \mathsf{y}_{a})}. 
\end{align}

\subsection{Gaussian-mixture expressions of the order-parameter functions}
Hereafter, we express the order-parameter functions as the following Gaussian mixtures: 
\begin{align}
Q_{\sigma}(\ket{\mathsf{x}})
&= q^{0}_{\sigma} \int dA_{Q} dH_{Q} \, q_{\sigma}(A_{Q}, H_{Q}) \left(\frac{\beta A_{Q}}{2\pi}\right)^{\frac{n}{2}}
\, \mathrm{e}^{-\frac{\beta A_{Q}}{2} \sum_{a} (\mathsf{x}_{a} - H_{Q})^{2}}, \label{GaussianMixtureQ}\\
\hat{Q}_{\sigma}(\ket{\mathsf{x}})
&= \hat{q}^{0}_{\sigma} \int d\hat{A}_{Q}d\hat{H}_{Q} \, \hat{q}_{\sigma}(\hat{A}_{Q}, \hat{H}_{Q}) \left(\frac{\beta \hat{A}_{Q}}{2\pi}\right)^{\frac{n}{2}} 
\, \mathrm{e}^{\frac{\beta \hat{A}_{Q}}{2} \sum_{a} (\mathsf{x}_{a} - \hat{H}_{Q})^{2}}, \label{GaussianMixtureQhat}\\
P_{\sigma \sigma^{\prime}}(\ket{\mathsf{y}})
&= p^{0}_{\sigma \sigma^{\prime}} \int dA_{P} dH_{P} \, p_{\sigma \sigma^{\prime}}(A_{P}, H_{P}) \left(\frac{\beta A_{P}}{2\pi}\right)^{\frac{n}{2}} 
\, \mathrm{e}^{-\frac{\beta A_{P}}{2} \sum_{a} (\mathsf{y}_{a} - H_{P})^{2}}, \label{GaussianMixtureP}\\
\hat{P}_{\sigma \sigma^{\prime}}(\ket{\mathsf{y}})
&= \hat{p}^{0}_{\sigma \sigma^{\prime}} \int d\hat{A}_{P}d\hat{H}_{P} \, \hat{p}_{\sigma \sigma^{\prime}}(\hat{A}_{P}, \hat{H}_{P}) \left(\frac{\beta \hat{A}_{P}}{2\pi}\right)^{\frac{n}{2}} 
\, \mathrm{e}^{\frac{\beta \hat{A}_{P}}{2} \sum_{a} (\mathsf{y}_{a} - \hat{H}_{P})^{2}} \label{GaussianMixturePhat}, 
\end{align}
where $q^{0}_{\sigma}$, $\hat{q}^{0}_{\sigma}$, $p^{0}_{\sigma \sigma^{\prime}}$, and $\hat{p}^{0}_{\sigma \sigma^{\prime}}$ are normalization factors. 

We first integrate with respect to $\hat{y}_{\alpha a}$. 
For a certain $\alpha \in V_{\sigma \sigma^{\prime}}$ ($\sigma \le \sigma^{\prime}$), 
\begin{align}
& \frac{1}{1+\delta(\sigma, \sigma^{\prime})} 
\int_{-\imgunit \infty}^{+\imgunit \infty} \prod_{a} \frac{d\hat{y}_{\alpha a}}{2\pi} \, \hat{P}_{\sigma \sigma^{\prime}} (\ket{\hat{y}}_{\alpha}) \hat{P}_{\sigma^{\prime} \sigma} (\ket{\hat{y}}_{\alpha}) 
\mathrm{e}^{-\beta \sum_{a} y_{\alpha a} \hat{y}_{\alpha a} }\notag\\
&= \frac{\hat{p}^{0}_{\sigma \sigma^{\prime}} \hat{p}^{0}_{\sigma^{\prime}\sigma}}{1+\delta(\sigma, \sigma^{\prime})}
\int d\hat{A}_{P}d\hat{H}_{P} d\hat{A}^{\prime}_{P}d\hat{H}^{\prime}_{P} 
\hat{p}_{\sigma\sigma^{\prime}}(\hat{A}_{P},\hat{H}_{P}) \hat{p}_{\sigma^{\prime}\sigma}(\hat{A}^{\prime}_{P},\hat{H}^{\prime}_{P}) \notag\\
&\hspace{10pt}\times 
\left( \frac{\beta \hat{A}_{P} \hat{A}^{\prime}_{P}}{ 2\pi(\hat{A}_{P} + \hat{A}^{\prime}_{P})} \right)^{\frac{n}{2}}
\exp\left( \frac{n\beta}{2}(\hat{A}_{P}\hat{H}^{2}_{P} + \hat{A}^{\prime}_{P}\hat{H}^{\prime 2}_{P}) 
- \frac{\beta}{2} \sum_{a} \frac{\left( \hat{A}_{P}\hat{H}_{P}+\hat{A}^{\prime}_{P}\hat{H}^{\prime}_{P}+y_{\alpha a} \right)^{2}}{\hat{A}_{P} + \hat{A}^{\prime}_{P}} 
 \right).
\end{align}

Then, we integrate with respect to $y_{\alpha a}$. 
For a certain $\alpha \in V^{0}$, irrespective of the group label, 
\begin{align}
&\int \prod_{a} dy_{\alpha a} 
\exp\left( 
\frac{\beta}{2} \sum_{a} y^{2}_{\alpha a} 
- \frac{\beta}{2} \frac{\sum_{a}(y_{\alpha a} + \hat{A}_{P}\hat{H}_{P} + \hat{A}^{\prime}_{P}\hat{H}^{\prime}_{P})^{2}}{\hat{A}_{P}+\hat{A}^{\prime}_{P}}
\right) \notag\\
&= \left( \frac{2\pi}{\beta} \frac{\hat{A}_{P} + \hat{A}^{\prime}_{P}}{1 - (\hat{A}_{P} + \hat{A}^{\prime}_{P})} \right)^{\frac{n}{2}} 
\exp\left( -\frac{n\beta}{2} \frac{(\hat{A}_{P}\hat{H}_{P} + \hat{A}^{\prime}_{P}\hat{H}^{\prime}_{P})^{2}}{\hat{A}_{P} + \hat{A}^{\prime}_{P}-1} 
\right).
\end{align}
Thus, for $\alpha \in V_{\sigma \sigma^{\prime}}$ ($\sigma \le \sigma^{\prime}$), 
\begin{align}
\Psi_{\alpha} 
&= \frac{\hat{p}^{0}_{\sigma \sigma^{\prime}} \hat{p}^{0}_{\sigma^{\prime}\sigma}}{1+\delta(\sigma, \sigma^{\prime})}
\int d\hat{A}d\hat{H} d\hat{A}^{\prime}d\hat{H}^{\prime} \, 
\hat{p}_{\sigma\sigma^{\prime}}(\hat{A}_{P},\hat{H}_{P}) \hat{p}_{\sigma^{\prime}\sigma}(\hat{A}^{\prime}_{P},\hat{H}^{\prime}_{P}) \notag\\
&\hspace{10pt}\times 
\left( \frac{\hat{A}_{P} \hat{A}^{\prime}_{P}}{1 - \hat{A}_{P} - \hat{A}^{\prime}_{P}} \right)^{\frac{n}{2}} 
\exp\left( \frac{n\beta}{2}(\hat{A}_{P}\hat{H}^{2}_{P} + \hat{A}^{\prime}_{P}\hat{H}^{\prime 2}_{P}) 
- \frac{n\beta}{2} \frac{(\hat{A}_{P}\hat{H}_{P} + \hat{A}^{\prime}_{P}\hat{H}^{\prime}_{P})^{2}}{\hat{A}_{P} + \hat{A}^{\prime}_{P}-1} 
\right).
\end{align}

We now calculate $\Phi_{i}$. 
By expanding $\mathrm{e}^{\hat{Q}_{\sigma}(\ket{x}_{i})}$ as 
\begin{align}
\mathrm{e}^{\hat{Q}_{\sigma}(\ket{x}_{i})} 
&= \sum_{d=0}^{\infty} \frac{1}{d!} \hat{Q}^{d}_{\sigma}(\ket{x}_{i}), 
\end{align}
we obtain 
\begin{align}
&\Phi_{i} = \sum_{d=0}^{\infty} \frac{1}{d!} 
\int \prod_{\ell=1}^{d}\left( \hat{q}^{0}_{\sigma} d\hat{A}_{Q \ell}d\hat{H}_{Q \ell} \, 
\left(\frac{\beta \hat{A}_{Q \ell}}{2\pi}\right)^{\frac{n}{2}}
\hat{q}_{\sigma}(\hat{A}_{Q \ell}, \hat{H}_{Q \ell}) \right)
\int \prod_{a} dx_{ia} \, \notag\\
&\hspace{20pt}\times\exp\left( 
- \beta \sum_{a=1}^{n} \sum_{r} h^{r}_{i} (\frac{\lambda}{2}  x^{2}_{ia} + \gamma x_{ia} - \hat{m}_{ra} x_{ia})
+ \frac{\beta}{2} \sum_{\ell} \hat{A}_{Q \ell} \sum_{a} (x_{ia} - \hat{H}_{Q \ell})^{2} 
\right) \notag\\
&= \sum_{d=0}^{\infty} \frac{\hat{q}^{0\, d}_{\sigma}}{d!} 
\int \prod_{\ell=1}^{d}\left( d\hat{A}_{Q \ell}d\hat{H}_{Q \ell} \, \hat{q}_{\sigma}(\hat{A}_{Q \ell}, \hat{H}_{Q \ell}) \right)
\left( \frac{\prod_{\ell} \hat{A}_{Q \ell} }{ \lambda \sum_{r} h^{r}_{i} - \sum_{\ell} \hat{A}_{Q \ell} } \right)^{\frac{n}{2}} \notag\\
&\hspace{10pt} \times\exp\left( 
\frac{n\beta}{2} \sum_{\ell} \hat{A}_{Q \ell} \hat{H}^{2}_{Q \ell} + \frac{\beta}{2} \frac{\sum_{a}\left( \sum_{r} h^{r}_{i}(\gamma - \hat{m}_{ra}) + \sum_{\ell} \hat{A}_{Q \ell}\hat{H}_{Q \ell} \right)^{2} }{ \lambda \sum_{r} h^{r}_{i} - \sum_{\ell} \hat{A}_{Q \ell} }
 \right).
\end{align}
We note that, in the limit where $n \to 0$, we have $\Phi_{i}(n=0) = \sum_{d=0}^{\infty} \frac{\hat{q}^{0\, d}_{\sigma}}{d!} = \mathrm{e}^{\hat{q}^{0}_{\sigma}}$, which yields the normalization factor of the Poisson distribution when the saddle point of $\log \Phi_{i}$ is taken.

To eliminate the apparent microscopic dependency on $\{h^{r}_{i}\}$, as defined in the main text, we introduce an empirical distribution: 
\begin{align}
\mathsf{P}_{\sigma}(\ket{h}) 
&= \frac{1}{N_{\sigma}} \sum_{i \in U_{\sigma}} 
\prod_{r} \delta\left(h^{r}, h^{r}_{i} \right), 
\end{align}
which yields 
\begin{align}
\frac{1}{N_{\sigma}} \sum_{i \in U_{\sigma}} \Phi_{i} 
&= \sum_{\ket{h}} \, \mathsf{P}_{\sigma}(\ket{h}) 
\sum_{d=0}^{\infty} \frac{\hat{q}^{0\, d}_{\sigma}}{d!} 
\int \prod_{\ell=1}^{d}\left( d\hat{A}_{Q \ell}d\hat{H}_{Q \ell} \, \hat{q}_{\sigma}(\hat{A}_{Q \ell}, \hat{H}_{Q \ell}) \right)
\left( \frac{2\pi}{\beta} \frac{1}{ \lambda \sum_{r} h^{r} - \sum_{\ell} \hat{A}_{Q \ell} } \right)^{\frac{n}{2}} \notag\\
&\hspace{10pt} \times\exp\left( 
\frac{n\beta}{2} \sum_{\ell} \hat{A}_{Q \ell} \hat{H}^{2}_{Q \ell} + \frac{n\beta}{2} \frac{\sum_{r} h^{r}\left( \gamma - \hat{m}_{r}) + \sum_{\ell} \hat{A}_{Q \ell}\hat{H}_{Q \ell} \right)^{2} }{ \lambda \sum_{r} h^{r} - \sum_{\ell} \hat{A}_{Q \ell} }
 \right). 
\end{align}

By using the Gaussian-mixture expressions, we can analogously calculate the integrals in $\Xi(n, \beta, \lambda, \gamma)$ as follows: 
\begin{align}
\int d\ket{\mathsf{x}} \, \hat{Q}_{\sigma}(\ket{\mathsf{x}}) Q_{\sigma}(\ket{\mathsf{x}}) 
&= q^{0}_{\sigma} \hat{q}^{0}_{\sigma} \int dA_{Q} dH_{Q} d\hat{A}_{Q} d\hat{H}_{Q} \, 
q_{\sigma}(A_{Q}, H_{Q}) \hat{q}_{\sigma}(\hat{A}_{Q}, \hat{H}_{Q}) 
\left( \frac{\beta A_{Q} \hat{A}_{Q}}{2\pi (A_{Q} - \hat{A}_{Q})} \right)^{\frac{n}{2}} \notag\\
&\hspace{20pt}\times \exp\left( -\frac{n\beta}{2}
\left(A_{Q}H^{2}_{Q} - \hat{A}_{Q}\hat{H}^{2}_{Q} - \frac{(A_{Q}H_{Q} - \hat{A}_{Q}\hat{H}_{Q})^{2}}{A_{Q} - \hat{A}_{Q}} \right) 
\right),
\end{align}

\begin{align}
\int d\ket{\mathsf{y}} \, \hat{P}_{\sigma \sigma^{\prime}}(\ket{\mathsf{y}}) P_{\sigma \sigma^{\prime}}(\ket{\mathsf{y}}) 
&= p^{0}_{\sigma \sigma^{\prime}} \hat{p}^{0}_{\sigma \sigma^{\prime}} 
\int dA_{P} dH_{P} d\hat{A}_{P}d\hat{H}_{P} \, 
p_{\sigma \sigma^{\prime}}(A_{P}, H_{P}) 
\hat{p}_{\sigma \sigma^{\prime}}(\hat{A}_{P}, \hat{H}_{P}) \notag\\
&\hspace{20pt}\times \left( \frac{\beta A_{P}\hat{A}_{P}}{ 2\pi(A_{P} - \hat{A}_{P})} \right)^{\frac{n}{2}} 
\exp\left( -\frac{n\beta}{2} \left( A_{P} H^{2}_{P} - \hat{A}_{P} \hat{H}^{2}_{P} \right) + \frac{n\beta}{2} \frac{(A_{P}H_{P} - \hat{A}_{P}\hat{H}_{P})^{2}}{A_{P}-\hat{A}_{P}} \right),
\end{align}

\begin{align}
&\int d\ket{\mathsf{x}} d\ket{\mathsf{y}} \, Q_{\sigma}(\ket{\mathsf{x}}) P_{\sigma \sigma^{\prime}}(\ket{\mathsf{y}}) 
\mathrm{e}^{-\beta \sum_{a} (\frac{\lambda}{2}  \mathsf{x}^{2}_{a} + \gamma \mathsf{x}_{a} - \mathsf{x}_{a} \mathsf{y}_{a})}
\notag\\
&= q^{0}_{\sigma} p^{0}_{\sigma \sigma^{\prime}} 
\int dA_{Q} dH_{Q} dA_{P}dH_{P} \, 
q_{\sigma}(A_{Q}, H_{Q}) 
p_{\sigma \sigma^{\prime}}(A_{P}, H_{P}) \left(\frac{\beta A_{Q}}{2\pi} \frac{\beta A_{P}}{2\pi}\right)^{\frac{n}{2}} \notag\\
&\hspace{10pt}\times \prod_{a=1}^{n}
\exp\left( 
-\frac{\beta}{2}\left( A_{Q}H^{2}_{Q} + A_{P}H^{2}_{P} \right) 
-\frac{\beta}{2} 
\begin{bmatrix}
\mathsf{x}_{a} \\
\mathsf{y}_{a}
\end{bmatrix}^{\top}
\begin{bmatrix}
\lambda + A_{Q} & -1 \\
-1 & A_{P}
\end{bmatrix}
\begin{bmatrix}
\mathsf{x}_{a} \\
\mathsf{y}_{a}
\end{bmatrix}
+ \beta 
\begin{bmatrix}
A_{Q}H_{Q} - \gamma \\
A_{P}H_{P}
\end{bmatrix}^{\top}
\begin{bmatrix}
\mathsf{x}_{a} \\
\mathsf{y}_{a}
\end{bmatrix}
\right) \notag\\
&= q^{0}_{\sigma} p^{0}_{\sigma \sigma^{\prime}} 
\int dA_{Q} dH_{Q} dA_{P}dH_{P} \, 
q_{\sigma}(A_{Q}, H_{Q}) 
p_{\sigma \sigma^{\prime}}(A_{P}, H_{P}) 
\left(\frac{A_{Q}A_{P}}{(\lambda + A_{Q})A_{P}-1}\right)^{\frac{n}{2}}
\notag\\
&\hspace{10pt}\times \exp\left( 
-\frac{n\beta}{2}\left( A_{Q}H^{2}_{Q} + A_{P}H^{2}_{P} \right) 
+\frac{n\beta}{2} 
\frac{A_{P} (A_{Q} H_{Q} - \gamma)^{2} + (\lambda + A_{Q}) A^{2}_{P} H^{2}_{P} + 2 A_{P}H_{P} (A_{Q}H_{Q} - \gamma) }{(\lambda + A_{Q}) A_{P}-1}
\right). 
\end{align}

\subsection{Normalization factors}
We now digress to calculate the normalization factors $q^{0}_{\sigma}$, $\hat{q}^{0}_{\sigma}$, $p^{0}_{\sigma \sigma^{\prime}}$, and $\hat{p}^{0}_{\sigma \sigma^{\prime}}$ in Eqs.~(\ref{GaussianMixtureQ})--(\ref{GaussianMixturePhat}). 
They can be derived from the estimate of $\log\left[ Z^{n} \right]_{B}$ with $n=0$. 
In the large-graph limit ($N \gg 1$), the saddle-point estimate of Eq.~(\ref{ZnMoment}) yields 
\begin{align}
& \log\left[ Z^{0} \right]_{B} = 0 \notag\\
&= \mathop{\mathop{\mathrm{extr}}_{\lambda, \gamma,\{q_{\sigma}\}, \{\hat{q}_{\sigma}\}, }}_{\{p_{\sigma\sigma^{\prime}}\}, \{\hat{p}_{\sigma\sigma^{\prime}}\}} 
\Biggl\{
\sum_{\sigma} \sum_{i \in U_{\sigma}} \log \Phi_{i}(n=0, \beta, \lambda, \gamma) 
+\sum_{\sigma \le \sigma^{\prime}} \sum_{\alpha \in V^{0}_{\sigma \sigma^{\prime}}} \log \Psi_{\alpha}(n=0, \beta) + \Xi(n=0, \beta)
\Biggr\} \notag\\
&= \mathop{\mathop{\mathrm{extr}}_{\{q^{0}_{\sigma}\}, \{\hat{q}^{0}_{\sigma}\}, }}_{\{p^{0}_{\sigma\sigma^{\prime}}\}, \{\hat{p}^{0}_{\sigma\sigma^{\prime}}\}} 
\Biggl\{
\sum_{\sigma} N_{\sigma} \hat{q}^{0}_{\sigma} 
+2\sum_{\sigma} e_{\sigma \sigma} \log \hat{p}^{0}_{\sigma \sigma}
+\sum_{\sigma < \sigma^{\prime}} e_{\sigma \sigma^{\prime}} \log(\hat{p}^{0}_{\sigma \sigma^{\prime}} \hat{p}^{0}_{\sigma^{\prime} \sigma}) \notag\\
&\hspace{60pt}- \sum_{\sigma} N_{\sigma} \hat{q}^{0}_{\sigma} q^{0}_{\sigma} 
- \sum_{\sigma, \sigma^{\prime}} e_{\sigma \sigma^{\prime}} \hat{p}^{0}_{\sigma \sigma^{\prime}} p^{0}_{\sigma \sigma^{\prime}} 
+ \sum_{\sigma, \sigma^{\prime}} N_{\sigma} e_{\sigma \sigma^{\prime}} q^{0}_{\sigma} p^{0}_{\sigma \sigma^{\prime}}
\Biggr\}. 
\end{align}
The saddle-point conditions of the equation above yield 
\begin{align}
&q^{0}_{\sigma} = 1, \\
&\hat{q}^{0}_{\sigma} = c_{\sigma}, \\
&\hat{p}^{0}_{\sigma \sigma^{\prime}} = N_{\sigma} \hspace{20pt} \text{ for any pair of $\sigma$ and $\sigma^{\prime}$}, \\
&p^{0}_{\sigma \sigma^{\prime}} = \frac{1+ \delta(\sigma, \sigma^{\prime})}{N_{\sigma}} \hspace{20pt} \text{ for any $\sigma$ and $\sigma^{\prime}$}. 
\end{align}
Here, we defined $c_{\sigma} \equiv N_{\sigma}^{-1} \sum_{\sigma^{\prime}} e_{\sigma \sigma^{\prime}} \left( 1 + \delta(\sigma, \sigma^{\prime}) \right)$ as in the main text.

\subsection{Saddle-point equations}
All the microscopic variables are now integrated out. 
Hereafter, we focus on the replica-symmetric solution with $m_{ra} = m_{r}$ and $\hat{m}_{ra} = \hat{m}_{r}$, i.e., we assume that there is no dependency on the replica indices. 
In the large-graph limit ($N \gg 1$), we evaluate $\log \left[ Z^{n}(\beta, \lambda, \gamma) \right]_{\mat{B}}$ using the saddle-point estimate. 
That is, 
\begin{align}
&\left[ \lambda_{2} \right]_{\mat{B}^{0}} 
= \mathop{\mathrm{extr}}_{\lambda, \gamma} \lim_{\beta\to\infty} \lim_{n \to 0} 
\frac{2}{\beta N} \frac{\partial}{\partial n} \log \left[ Z^{n}(\beta, \lambda, \gamma) \right]_{\mat{B}^{0}} \notag\\
&\mathop{\rightarrow}_{N \gg 1} \, 
\mathop{\mathop{\mathrm{extr}}_{\lambda, \gamma, \{m_{r}\}, \{\hat{m}_{r}\},}}_{\{q_{\sigma}\}, \{\hat{q}_{\sigma}\}, \{p_{\sigma\sigma^{\prime}}\}, \{\hat{p}_{\sigma\sigma^{\prime}}\}} 
\Biggl\{
\frac{\lambda}{2}
+ \sum_{r=1}^{R} \left( \frac{N}{d^{v}_{r}} m_{r}^{2} - \hat{m}_{r} m_{r} \right) \notag\\
&\hspace{80pt}+ \lim_{\beta \to\infty} \lim_{n \to 0} \frac{1}{\beta N} \frac{\partial}{\partial n} \sum_{\sigma} \sum_{i \in U_{\sigma}} \log \Phi_{i}(n, \beta, \lambda, \gamma, \{\hat{m}_{r}\}) \notag\\
&\hspace{80pt}+ \lim_{\beta \to\infty} \lim_{n \to 0} \frac{1}{\beta N} \frac{\partial}{\partial n} \sum_{\sigma \le \sigma^{\prime}} \sum_{\alpha \in V_{\sigma \sigma^{\prime}}} \log \Psi_{\alpha}(n, \beta) 
+ \lim_{\beta \to\infty} \lim_{n \to 0} \frac{1}{\beta N} \frac{\partial}{\partial n} \Xi(n, \beta, \lambda, \gamma)
\Biggr\}. \label{SaddlePointEq}
\end{align}
Here, we assumed that the extremization and the limits with respect to $\beta$ and $n$ can be interchanged.

The saddle-point conditions yield the following message-passing equations with respect to the mean and variance of the Gaussian mixtures in the order-parameter functions: 
\begin{align}
q_{\sigma}\left( A_{Q}, H_{Q} \right) 
&= \sum_{\ket{h}} \mathsf{P}_{\sigma}(\ket{h}) 
\sum_{d=0}^{\infty} \mathcal{P}_{c_{\sigma}}\left(d\right) 
\int\prod_{\ell=1}^{d}\left( d\hat{A}_{Q \ell}d\hat{H}_{Q \ell} \, \hat{q}_{\sigma}\left( \hat{A}_{Q \ell}, \hat{H}_{Q \ell} \right) \right) \notag\\
&\hspace{20pt}\times \delta\left( A_{Q} + \sum_{\ell=1}^{d} \hat{A}_{Q \ell} - \lambda \sum_{r} h^{r} \right) 
\delta\left( H_{Q} - \frac{ \sum_{r} h^{r}(\gamma - \hat{m}_{r}) + \sum_{\ell=1}^{d}\hat{A}_{Q \ell}\hat{H}_{Q \ell} }{\sum_{\ell=1}^{d} \hat{A}_{Q \ell} - \lambda \sum_{r} h^{r} } \right), \\
\hat{q}_{\sigma}\left( \hat{A}_{Q}, \hat{H}_{Q} \right) 
&= \sum_{\sigma^{\prime}} \frac{e_{\sigma \sigma^{\prime}} \left( 1+\delta(\sigma, \sigma^{\prime}) \right)}{c_{\sigma}N_{\sigma}} \notag\\
&\hspace{20pt} \times \int dA_{P}dH_{P} \, p_{\sigma \sigma^{\prime}}\left(A_{P}, H_{P}\right) 
\delta\left( \hat{A}_{Q} - \frac{1 - \lambda A_{P}}{A_{P}} \right) 
\delta\left( \hat{H}_{Q} - \frac{A_{P}\left( \gamma - H_{P} \right)}{1 - \lambda A_{P}} \right), \\
p_{\sigma \sigma^{\prime}}\left( A_{P}, H_{P} \right) 
&= \int d\hat{A}_{P}d\hat{H}_{P} \, \hat{p}_{\sigma^{\prime} \sigma}\left( \hat{A}_{P}, \hat{H}_{P} \right) 
\delta\left( A_{P} - 1 + \hat{A}_{P} \right) 
\delta\left( H_{P} + \frac{\hat{A}_{P}\hat{H}_{P}}{1 - \hat{A}_{P}} \right), \\
\hat{p}_{\sigma \sigma^{\prime}}\left( \hat{A}_{P}, \hat{H}_{P} \right) 
&= \int dA_{Q}dH_{Q} \, q_{\sigma}\left( A_{Q}, H_{Q} \right) 
\delta\left( \hat{A}_{P} - \frac{1}{\lambda + A_{Q}} \right)
\delta\left( \hat{H}_{P} - \gamma + A_{Q}H_{Q} \right),
\end{align}
where 
\begin{align}
\mathcal{P}_{c}\left(d\right) \equiv \frac{c^{d}}{d!} \mathrm{e}^{-c}
\end{align}
is a Poisson distribution with mean $c$ and represents the degree distribution of the physical nodes.

By combining the above equations, we arrive at 
\begin{align}
q_{\sigma}\left( A_{Q}, H_{Q} \right) 
&= \sum_{\ket{h}} \mathsf{P}_{\sigma}(\ket{h}) 
\sum_{d=0}^{\infty} \mathcal{P}_{c_{\sigma}}\left(d\right) 
\prod_{\ell=1}^{d}\left( \sum_{\sigma^{\prime}} f_{\sigma \sigma^{\prime}} 
\int dA_{Q\ell}dH_{Q\ell} \, q_{\sigma^{\prime}}\left( A_{Q\ell}, H_{Q\ell} \right) \right)
\notag\\
&\hspace{20pt}\times \delta\left( A_{Q} - \lambda (d + \sum_{r} h^{r}) + \sum_{\ell=1}^{d} \left( 1 - \frac{1}{\lambda + A_{Q\ell}} \right)^{-1} \right) \notag\\
&\hspace{20pt}\times \delta\left( H_{Q} + \frac{ \gamma(d + \sum_{r}h^{r}) - \left(\sum_{r} h^{r}\hat{m}_{r} + \sum_{\ell=1}^{d} \left( \frac{\gamma - A_{Q\ell}H_{Q\ell}}{1-(\lambda + A_{Q\ell})} \right) \right) }{\lambda (d + \sum_{r} h^{r}) - \sum_{\ell=1}^{d} \left( 1 - \frac{1}{\lambda + A_{Q\ell}} \right)^{-1}} \right), \label{PhysicalCavityEq}
\end{align}
which is the self-consistent equation in the main text, where $A_{Q}$ and $H_{Q}$ are replaced with $\mathsf{A}$ and $\mathsf{H}$, respectively.

From the other saddle-point conditions, we obtain 
\begin{align}
& m_{r} = \frac{d^{v}_{r}}{2 N} \hat{m}_{r}, \label{SaddlePoint-mr}\\
& m_{r} = \sum_{\sigma} \frac{N_{\sigma}}{N} \sum_{d=0}^{\infty} \mathcal{P}_{c_{\sigma}}(d) 
\sum_{\ket{h}} \mathsf{P}_{\sigma}(\ket{h}) \, h^{r}  
\int \prod_{\ell=1}^{d} \left( d\hat{A}_{Q \ell}d\hat{H}_{Q \ell} \, \hat{q}_{\sigma}\left( \hat{A}_{Q \ell}, \hat{H}_{Q \ell} \right) \right) 
\frac{ \sum_{r} h^{r}(\gamma - \hat{m}_{r}) + \sum_{\ell=1}^{d}\hat{A}_{Q \ell}\hat{H}_{Q \ell} }{\sum_{\ell=1}^{d} \hat{A}_{Q \ell} - \lambda \sum_{r} h^{r} }, \label{SaddlePoint-hatmr}\\
& \hat{m}_{r} = \frac{1}{N} \sum_{\sigma \sigma^{\prime}} \left( 1 + \delta(\sigma, \sigma^{\prime}) \right) e_{\sigma \sigma^{\prime}} 
\int dA_{Q}dH_{Q} \, q_{\sigma}\left( A_{Q}, H_{Q} \right) 
\int dA_{P}dH_{P} \, p_{\sigma \sigma^{\prime}}\left(A_{P}, H_{P}\right) \, 
\frac{A_{P}(A_{Q}H_{Q} + H_{P} - \gamma)}{1 - A_{P}(\lambda + A_{Q})}, \label{SaddlePoint-gamma}\\
& 1 = \sum_{\sigma} \frac{N_{\sigma}}{N} \sum_{d=0}^{\infty} \mathcal{P}_{c_{\sigma}}(d) 
\sum_{\ket{h}} \mathsf{P}_{\sigma}(\ket{h}) 
\sum_{r} h^{r} \notag\\
& \hspace{20pt}\times \int \prod_{\ell=1}^{d} \left( d\hat{A}_{Q \ell}d\hat{H}_{Q \ell} \, \hat{q}_{\sigma}\left( \hat{A}_{Q \ell}, \hat{H}_{Q \ell} \right) \right) 
\left( \frac{ \sum_{r} h^{r}(\gamma - \hat{m}_{r}) + \sum_{\ell=1}^{d}\hat{A}_{Q \ell}\hat{H}_{Q \ell} }{\sum_{\ell=1}^{d} \hat{A}_{Q \ell} - \lambda \sum_{r} h^{r} } \right)^{2} \notag\\
& \hspace{20pt} + \frac{1}{N} \sum_{\sigma \sigma^{\prime}} \left( 1 + \delta(\sigma, \sigma^{\prime}) \right) e_{\sigma \sigma^{\prime}} 
\int dA_{Q}dH_{Q} \, q_{\sigma}\left( A_{Q}, H_{Q} \right) 
\int dA_{P}dH_{P} \, p_{\sigma \sigma^{\prime}}\left(A_{P}, H_{P}\right) 
\left( \frac{A_{P}(A_{Q}H_{Q} + H_{P} - \gamma)}{1 - A_{P}(\lambda + A_{Q})} \right)^{2}. \label{SaddlePoint-lambda}
\end{align}
Equations (\ref{SaddlePoint-hatmr})--(\ref{SaddlePoint-lambda}) are the self-consistent equations for $m_{r}$, $\lambda$, and $\gamma$. 

\end{widetext}

\section{Relation to the crude approximation: Small-fluctuation limit}\label{sec:SmallFluctuationLimit}
In this section, we consider the limit at which the variance of the eigenvector-element distributions is negligibly small, i.e., the distribution of the precision parameter $A_{Q}$ has a peak at an infinitely large value. 
In this case, the eigenvector element in group $\sigma$ can be well characterized by 
\begin{align}
\bracket{H}_{\sigma} 
&\equiv \int dA_{Q}dH_{Q} q_{\sigma}\left( A_{Q}, H_{Q} \right) \, H_{Q}, \label{Haverage1}
\end{align}
which corresponds to $\bar{\varphi}_{2 \sigma}$ in the crude approximation. 
In the following, in the absence of external hyperedges, we show that the eigenvalue equation under the crude approximation can indeed be recovered. 
We also show that the saddle-point equations (\ref{SaddlePoint-gamma}) and (\ref{SaddlePoint-lambda}) represent the orthogonality and normalization conditions that appear as the constraints in the original optimization problem. 

\subsection{Mean eigenvalue equation}
To derive the equation of the small-fluctuation limit, we first assume that the precision parameter $A_{Q}$ can be represented by a single number $a$, irrespective of specific node labels or group labels; this is known as the effective medium approximation.  
Equation (\ref{Haverage1}) is calculated as follows: 
\begin{widetext}
\begin{align}
\bracket{H}_{\sigma} 
&= \sum_{\ket{h}} \mathsf{P}_{\sigma}(\ket{h}) 
\sum_{d=0}^{\infty} \mathcal{P}_{c_{\sigma}}\left(d\right) 
\prod_{\ell=1}^{d}\left( \sum_{\sigma^{\prime}} \frac{e_{\sigma \sigma^{\prime}}
\left( 1+\delta(\sigma, \sigma^{\prime}) \right)}{c_{\sigma} N_{\sigma}} 
\int dA_{Q\ell}dH_{Q\ell} \, q_{\sigma^{\prime}}\left( A_{Q\ell}, H_{Q\ell} \right) \right) \notag\\
&\hspace{20pt}\times \int dA_{Q}dH_{Q} \delta\left( A_{Q} - \lambda (d + \sum_{r} h^{r}) + \sum_{\ell=1}^{d} \left( 1 - \frac{1}{\lambda + A_{Q\ell}} \right)^{-1} \right) \notag\\
&\hspace{20pt}\times H_{Q} \, \delta\left( H_{Q} + \frac{ \gamma(d + \sum_{r}h^{r}) - \left(\sum_{r} h^{r}\hat{m}_{r} + \sum_{\ell=1}^{d} \left( \frac{\gamma - A_{Q\ell}H_{Q\ell}}{1-(\lambda + A_{Q\ell})} \right) \right) }{\lambda (d + \sum_{r} h^{r}) - \sum_{\ell=1}^{d} \left( 1 - \frac{1}{\lambda + A_{Q\ell}} \right)^{-1}} \right)\notag\\
&= \sum_{\ket{h}} \mathsf{P}_{\sigma}(\ket{h}) 
\sum_{d=0}^{\infty} \mathcal{P}_{c_{\sigma}}\left(d\right) 
\left( \lambda \sum_{r}h^{r} + d\left( \lambda - \frac{\lambda + a}{\lambda -1 + a} \right) \right)^{-1} 
\left( 2 N \sum_{r} \frac{h^{r} m_{r}}{d^{v}_{r}} + \frac{d a}{\lambda - 1 + a} \sum_{\sigma^{\prime}} f_{\sigma \sigma^{\prime}} \bracket{H}_{\sigma^{\prime}} \right).
\end{align}
Here, we used the fact that $\gamma = 0$. 
Taking the limit where $a \to \infty$, we have 
\begin{align}
\bracket{H}_{\sigma} 
&= \sum_{\ket{h}} \mathsf{P}_{\sigma}(\ket{h}) 
\sum_{d=0}^{\infty} \mathcal{P}_{c_{\sigma}}\left(d\right) 
\frac{ 2 N \sum_{r} h^{r} m_{r} /d^{v}_{r} + d \sum_{\sigma^{\prime}} f_{\sigma \sigma^{\prime}} \bracket{H}_{\sigma^{\prime}} }{ \lambda \sum_{r}h^{r} + d\left( \lambda - 1 \right) }, \label{SmallFluctEqn}
\end{align}
which is an improved version of the crude approximation.
In the absence of external hyperedges, it is confirmed that the equation under the crude approximation (Eq.~(\ref{CrudeApproxUnperturbed})) is recovered. 
\begin{align}
\sum_{\sigma^{\prime}} f_{\sigma \sigma^{\prime}} \bracket{H}_{\sigma^{\prime}} 
= \left( \lambda - 1 \right) \bracket{H}_{\sigma}. 
\end{align}
\end{widetext}

\subsection{Orthogonality and normalization conditions}
The saddle-point equations (\ref{SaddlePoint-gamma}) and (\ref{SaddlePoint-lambda}) are evidently highly complicated. 
In fact, Eq.~(\ref{SaddlePoint-gamma}) represents the orthogonality condition $\sum_{i \in U} d^{u}_{i} x_{i} = 0$, whereas Eq.~(\ref{SaddlePoint-lambda}) represents the normalization condition $\sum_{i \in U} d^{u}_{i} x^{2}_{i} = N$. 
They are not easily comprehensible, because the degree $d^{u}_{i}$ of a physical node depends on the distribution of the external hyperedges as well as the distribution of the incidence matrix $\mat{B}^{0}$ of the original graph. 
Here, we show that they indeed represent the orthogonality and normalization conditions in the absence of external hyperedges in the small-fluctuation limit. 

In the absence of the external hyperedges, it is apparent from Eq.~(\ref{SaddlePoint-hatmr}) that $m_{r} = 0$. 
Thus, the left-hand side of Eq.~(\ref{SaddlePoint-gamma}) is zero by Eq.~(\ref{SaddlePoint-mr}). 
For the right-hand side of Eq.~(\ref{SaddlePoint-gamma}), by substituting the message-passing equations with respect to $p_{\sigma \sigma^{\prime}}$ and $\hat{p}_{\sigma \sigma^{\prime}}$, we obtain 
\begin{align}
& \sum_{\sigma, \sigma^{\prime}} \frac{c_{\sigma} N_{\sigma}}{N} f_{\sigma \sigma^{\prime}} 
\int dA_{Q}dH_{Q} \, q_{\sigma}\left( A_{Q}, H_{Q} \right) \notag\\
&\hspace{50pt}\times \int dA^{\prime}_{Q}dH^{\prime}_{Q} \, q_{\sigma^{\prime}}\left( A^{\prime}_{Q}, H^{\prime}_{Q} \right) \notag\\
&\hspace{50pt}\times \frac{A_{Q}H_{Q} - A^{\prime}_{Q}H^{\prime}_{Q} + (\lambda + A^{\prime}_{Q})(\gamma - A_{Q}H_{Q}) }{A_{Q} - A^{\prime}_{Q} + (\lambda + A_{Q}) (\lambda + A^{\prime}_{Q})}. 
\end{align}
In the small-fluctuation limit, only the second-order term in the numerator of the integrand remains. 
Thus, Eq.~(\ref{SaddlePoint-gamma}) becomes 
\begin{align}
0 = \sum_{\sigma} \frac{c_{\sigma} N_{\sigma}}{N} \bracket{H}_{\sigma}. 
\end{align}
Similarly, the first term in Eq.~(\ref{SaddlePoint-lambda}) is zero in the absence of external hyperedges. 
Thus, 
\begin{align}
1 = \sum_{\sigma} \frac{c_{\sigma} N_{\sigma}}{N} \bracket{H^{2}}_{\sigma}, 
\end{align}
where $\bracket{H^{2}}_{\sigma} \equiv \int dA_{Q}dH_{Q} q_{\sigma}\left( A_{Q}, H_{Q} \right) \, H^{2}_{Q}$. 
This indicates that the orthogonality and normalization constraints of the eigenvector elements $\{ \phi_{2i} \}$ are expressed by the distribution of $H_{Q}$.

\section{NMF on scotch-taped graphs}\label{sec:NMF}
Herein, we briefly discuss the application of NMF to scotch-taped graphs. 
We use the implementation of the NMF in \textit{scikit-learn} \cite{sklearn-NMF}. 

\begin{figure}[t!]
  \centering
  \includegraphics[width= 0.9\columnwidth, bb=0 0 472 742]{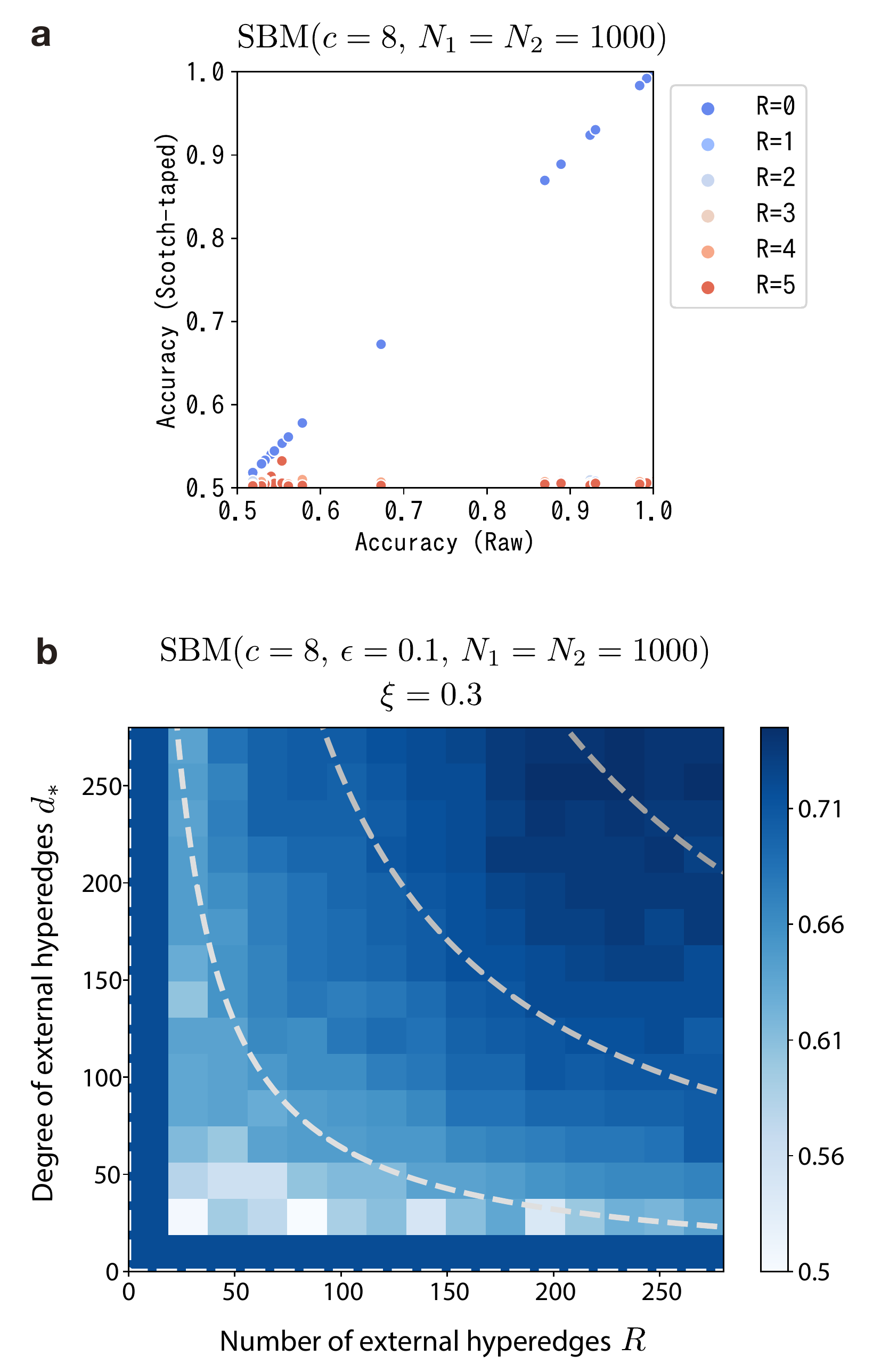}
  \caption{
  Numerical experiments of the NMF on the symmetric SBM ($c=8$, $N_{1}=N_{2}=1000$) corresponding to ({\bf a}) Fig.~\ref{fig:OverlapComparisons}{\bf b} and ({\bf b}) Fig.~\ref{fig:OverlapDensityPlot}{\bf b} in the main text. 
  In ({\bf a}), we consider various values of $\epsilon$ ranging from $\epsilon = 0.01$ to $0.2$.   
  }
  \label{fig:NMF}
\end{figure}

We conduct the same type of experiments as described in Sec.~\ref{AnalysisMessagePassing}. 
Figure \ref{fig:NMF}{\bf a} shows the case of uniform external hyperedges, corresponding to Fig.~\ref{fig:OverlapComparisons}{\bf b} in Sec.~\ref{sec:Type1Analysis}.
An experiment corresponding to Fig.~\ref{fig:OverlapDensityPlot}{\bf b} in Sec.~\ref{sec:R-degreeDependence} is shown in Fig.~\ref{fig:NMF}{\bf b}. 
In these experiments, we generated symmetric SBM instances with $c=8$, $N_{1} = N_{2} = 1000$. 
(Although we could consider $c=12$, we selected $c=8$ because the behavior of the NMF can be better observed with $c=8$.) 
It is evident from these results that, even a few the external hyperedges considerably modify the module structure inferred using the original graph.

\bibliography{ref}

\begin{thebibliography}{31}
\bibitem [{Com()}]{CommentLippmannSchwinger}%
  \BibitemOpen
  \href@noop {} {}\bibinfo {howpublished} {This is not exactly the
  Lippmann--Schwinger equation because we included the eigenvalue after the
  perturbation (i.e., scotch-taping) in the perturbation term.}\BibitemShut
  {Stop}%
\end{thebibliography}

\end{document}